\documentclass[preprint,12pt]{elsarticle}

\usepackage{amssymb}

\usepackage{booktabs}
\usepackage{makecell}
\usepackage{graphicx}
\usepackage{natbib}
\usepackage{algorithm}
\usepackage{algpseudocode}
\usepackage[all,cmtip]{xy}
\usepackage{tikz}
\usetikzlibrary{matrix}
\usepackage{tikz-cd}

\usepackage{amssymb}
\usepackage{epstopdf}
\usepackage{natbib}
\usepackage{amsmath, mathabx}
\usepackage{mathrsfs}
\usepackage{hyperref}
\usepackage{comment}
\usepackage{xcolor}
\usepackage{subfigure}
\usepackage{enumitem}
\usepackage{algorithm}
\usepackage{algpseudocode}
\usepackage{stmaryrd}
\usepackage[utf8]{inputenc}
\usepackage[T1]{fontenc}
\usepackage{fancyhdr}

\usepackage{cleveref}

\allowdisplaybreaks

 \hypersetup{
     colorlinks=true,
     citecolor=blue,
     linkcolor=blue,
     filecolor=magenta,      
     urlcolor=cyan,
 }

\DeclareMathAlphabet\mathbfcal{OMS}{cmsy}{b}{n}

\newcommand{\cE}{\mathcal{E}}

\newcommand{\cM}{\mathcal{M}}

\newcommand{\cS}{\mathcal{S}}

\newcommand{\cX}{\mathcal{X}}
\newcommand{\cY}{\mathcal{Y}}

\journal{arXiv}

\begin{document}

\begin{frontmatter}



\title{Cocaine Use Prediction with Tensor-based Machine Learning on Multimodal MRI Connectome Data}


\author[label1,label4]{Anru R. Zhang\fnref{1}}
\author[label2]{Ryan P. Bell\fnref{1}}
\author[label3]{Chen An\fnref{1}}
\author[label5]{Runshi Tang\fnref{1}}
\author[label2]{Shana A. Hall}
\author[label1]{Cliburn Chan}
\author[label2]{Kareem Al-Khalil}
\author[label2]{Christina S. Meade\corref{cor1}}

\cortext[cor1]{Corresponding author. Email address: \ead{christina.meade@duke.edu}}
\fntext[1]{These authors have equally contributed to this paper.}

\affiliation[label1]{organization={Department of Biostatistics and Bioinformatics, Duke Univeristy},
            city={Durham},
            postcode={27710}, 
            state={NC},
            country={U.S.}}

\affiliation[label2]{organization={Department of Psychiatry and Behavioral Sciences, Duke Univeristy},
	city={Durham},
	postcode={27710}, 
	state={NC},
	country={U.S.}}

\affiliation[label3]{organization={Department of Mathematics, Duke Univeristy},
	city={Durham},
	postcode={27708}, 
	state={NC},
	country={U.S.}}

\affiliation[label4]{organization={Department of Computer Science, Duke University},
	city={Durham},
	postcode={27710}, 
	state={NC},
	country={U.S.}}

\affiliation[label5]{organization={Department of Statistics, University of Wisconsin–Madison},
	city={Madison},
	postcode={53706}, 
	state={WI},
	country={U.S.}}

\begin{abstract}
This paper considers the use of machine learning algorithms for predicting cocaine use based on magnetic resonance imaging (MRI) connectomic data. The study utilized  functional MRI (fMRI) and diffusion MRI (dMRI) data collected  from 275 individuals, which was then parcellated into 246 regions of interest (ROIs) using the Brainnetome atlas. After data preprocessing, the datasets were transformed into tensor form. We developed a tensor-based unsupervised machine learning algorithm to reduce the size of the data tensor from $275$ (individuals) $\times 2$ (fMRI and dMRI) $\times 246$ (ROIs) $\times 246$ (ROIs) to $275$ (individuals) $\times 2$ (fMRI and dMRI) $\times 6$ (clusters) $\times 6$ (clusters). This was achieved by applying the high-order Lloyd algorithm to group the ROI data into 6 clusters. Features were extracted from the reduced tensor and combined with demographic features (age, gender, race, and HIV status). The resulting dataset was used to train a Catboost model using subsampling and nested cross-validation techniques, which achieved a prediction accuracy of 0.857 for identifying cocaine users. The model was also compared with other models, and the feature importance of the model was presented.

Overall, this study highlights the potential for using tensor-based machine learning algorithms to predict cocaine use based on MRI connectomic data and presents a promising approach for identifying individuals at risk of substance abuse. 
\end{abstract}




\end{frontmatter}

\begin{sloppypar}
\section{Introduction}\label{sec:intro}

Worldwide, cocaine was used by 20 million (0.4\%) individuals between the ages of 15-64 in 2019 (United Nations Office on Drugs and Crime, 2021), and an estimated 2\% of persons aged 12 or older in the United States used cocaine in the past year \citep{substance_abuse}. Chronic cocaine use is associated with both structural and functional deficits in the brain. Individuals who use cocaine show lower white matter integrity consistently within the corpus callosum and frontal regions \citep{suchting2021meta,beard2019regional}, and also across  diffuse association and projection fibers \cite{gaudreault2022whole}, as measured by diffusion tensor imaging (DTI), a method of examining white matter microstructure in magnetic resonance imaging (MRI) data. Resting-state functional MRI (fMRI) studies have been utilized to identify alterations in cerebral blood flow within and between functional networks in people who use cocaine. These alterations include deficits in functional connectivity within and between neural networks associated with reward processing and executive functioning \citep{berlingeri2017resting,hobkirk2019reward,hu2015impaired}. Research shows altered reorganization of functional connectomes involving executive, reward, salience, and default mode networks \citep{yip2019connectome,konova2015effects,liang2015interactions}. Other studies integrating structural and functional connectomics show cocaine-related effects involving the interoceptive networks \citep{de2019multimodal}. 

Connectomics is a method of describing brain networks based on data collected via non-invasive MRI techniques, including fMRI and diffusion MRI (dMRI) \cite{sporns2013human}. MRI connectome data provides a powerful way to map  cortical and subcortical connectivity in the brain \cite{BH16}. Brain connectomic data can be used to identify indicators of structural and functional neural deficits \cite{BB09}. Even though MRI investigations have resulted in consistent findings of structural and functional differences related to cocaine use, they typically employ unimodal analyses. Recent theories on brain disease envision the interplay of multiple neural regions, as captured by multiple MRI modalities, contributing to neurological deficits \cite{calhoun2016multimodal, liu2015multimodal, mandal2023brain}. There is also a surge of interest in investigating human brain connectomes using multiple imaging modalities and exploring the association with phenotypic traits. In other brain disorders, machine learning methods have been applied to MRI connectome data. For example, connectome features have been used to identify patients with schizophrenia, major depressive disorder, attention deficit hyperactivity disorder, and bipolar disorder \cite{BH16}.

\begin{figure}[h]
	\begin{center}
		\includegraphics[scale=0.55]{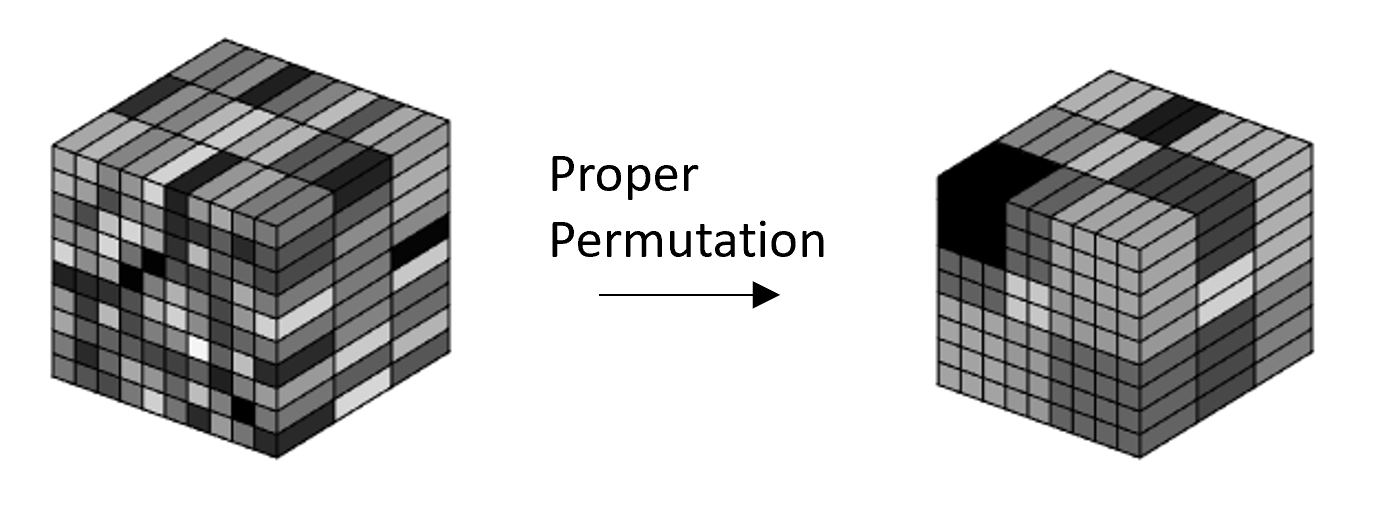}
		\caption{Demonstartion of a blockwise tensor}
		\label{fig_blockwise_tensor}
	\end{center}
\end{figure}

The aim of this paper is to develop a novel tensor-based machine learning method for predicting cocaine use based on multimodal MRI. Tensors are a high-order generalization of matrices, and they have various applications in different fields, including social networks, computer vision, neuroscience, and genomics. Tensors provide an effective representation of the hidden structure in multiway data, and low rankness is one of the popular structures utilized in tensor data analysis. If we partition each direction of a tensor, also known as a tensor mode, into multiple clusters where values within the same cluster are similar, we refer to this tensor as a blockwise tensor. Upon proper permutation of its indices, a blockwise tensor reveals blockwise structures \cite{Han20}. An illustrative example can be seen in Figure \ref{fig_blockwise_tensor}. Extracting and representing this blockwise information in a smaller-sized tensor enables dimension reduction. The tensor-based approach offers a significant advantage by preserving the original structural information of fMRI and dMRI connectome matrices independently, without the need for vectorization.

Our approach involves two main steps: first, a tensor-based unsupervised machine learning technique called high-order Lloyd \cite{Han20} is employed to reduce the dimensionality of the multimodal connectome data. This unsupervised machine learning approach distinguishes itself from previous work by utilizing the tensorial structure of the correlations between specific regions of interest (ROIs) based on an a priori brain atlas as the feature space. Second, a supervised machine learning algorithm is applied to predict cocaine use status using both the dimension-reduced diffusion and functional connectomes, along with demographic variables, as inputs to the model. Figure \ref{outline} provides an overview of the methodology, and Figure \ref{fig:procedure} illustrates each step of the procedure. The study demonstrates the feasibility of accurately predicting cocaine use status using a tensor-based unsupervised machine learning paradigm.

The method developed in this paper results in more accurate and/or efficient prediction in distinguishing cocaine users from non-users than other supervised and unsupervised machine learning paradigms in the sample. We explored several classical approaches as part of our study, such as training a classifier without dimension reduction, preprocessing with Principal Component Analysis (PCA), and clustering based on Brain atlas information. However, these methods yielded lower accuracy compared to our proposed approach. Moreover, we referred to a similar study by \cite{gowin2019using}, where they applied other supervised learning algorithms (including linear models and random forests) directly on a separate dataset without dimension reduction. Their results also showed lower accuracy. An important advantage of our method is its high interpretability, efficiency, and predictability. Meaningful partitions and high prediction performance were found, demonstrating the potential of the tensor-based approach for predicting cocaine use status based on diffusion and functional brain connectivity data.

\section{Methods}\label{sec:data}

\subsection{Sampling}\label{sec:sampling}

This is a secondary data analysis of MRI studies that examined the effects of HIV disease and substance use disorders on brain structure and function. The eligibility criteria are similar to those in \cite{Hetal21}. In brief, the study was open to adults aged 18-60 with or without a history of cocaine use. Cocaine use was operationalized as any regular cocaine use for $\ge 1$ year and either $\ge 2$ days of use in the past $30$ days. Non-cocaine use was operationalized as: no lifetime cocaine use (abuse or dependence), no history of regular cocaine use, no cocaine use in the past year, and a cocaine-negative urine drug screen. Alcohol, marijuana, and nicotine use were permitted in all groups. For other substances, including methamphetamine, heroin, and controlled medications that were not prescribed, participants were excluded for any lifetime dependence, regular use for $>2$ years, and regular use in the past $90$ days. Additional exclusion criteria were: non-fluency in English; illiteracy; less than 8th-grade education; severe learning disability; severe mental illness; current use of antipsychotic or mood-stabilizing medications; serious neurological disorders or history of neuro-infections; severe head trauma with loss of consciousness $>30$ min and persistent functional decline; MRI contraindications; and/or impaired mental status (i.e., not acutely psychotic, manic, delirious, or intoxicated).

\subsection{Procedures}\label{sec:procedures}

We recruited participants through advertisements in local newspapers, websites, community-based organizations, and healthcare clinics. After a short pre-screener to identify individuals with obvious exclusion criteria, individuals completed an in-person screening. HIV status was determined using an OraQuick rapid test. Participants who self-reported that they were diagnosed with HIV could skip the rapid test, and the diagnosis was verified using healthcare records. Eligible participants returned for the MRI scan and additional assessments. All participants provided written informed consent, and procedures were approved by the institutional review board at the Duke University Health System. Participants were compensated up to \$180 for their participation.

All participants had a confirmed blood alcohol level of $0.00$ before completing the MRI. Participants completed clinical interviews, computerized surveys, urine drug screening, and pregnancy testing. Module E of the Structured Clinical Interview for DSM-IV-TR identified substance use disorders, psychotic disorders, and bipolar disorder \cite{FSGW96}, and the Addiction Severity Index-Lite assessed current and lifetime substance use and associated impairments \cite{TMC06}. A medical history interview was used to assess the history of neurologic disorders (e.g. stroke). Healthcare records were reviewed to ensure no exclusionary medical history, including substance abuse. Participants reported demographic characteristics, including age, gender, race, and education.

\subsection{MRI data acquisition}\label{sec:MRI-acquisition}

MRI data were combined across five MRI protocols. The data were acquired on a 3.0T GE Discovery MR750 whole-body scanner with an 8-channel head coil. High-resolution T1-weighted (T1w) images were obtained using a spoiled echo sequence (repetition time [TR] = 8.16ms or 8.10ms, echo time [TE] = 3.18ms, voxel size = 1mm $\times$ 1mm $\times$ 1mm, field of view [FOV] = 256mm$^2$, 256-by-256 matrix, 12$^{\circ}$ flip angle, all protocols had either 166 or 168 interleaved slice data acquisition).

Diffusion-weighted imaging (DWI) data were acquired in the axial plane using a diffusion-sensitized parallel echo-planar sequence (FOV= 256 mm$^2$, voxel size = 2mm $\times$ 2mm $\times$ 2mm, 128-by-128 matrix, 90$^{\circ}$ flip angle, interleaved slices). To maximize the signal-to-noise ratio, the TE was set to use the minimum. Several parameters differed slightly between protocol 1 (b-factor = 900 s/mm$^2$, TR/TE = 10,000/81.5-84.6, 73 slices), protocol 2 (b-factor= 800 s/mm$^2$, TR/TE= 8000/77.9-83.9, 67 slices), protocol 3 (b-factor=800 s/mm$^2$, TR/TE = 8000/77.9-81.0, 67 slices), protocol 4 (b-factor = 800 s/mm$^2$, TR/TE = 8000/77.9-78.9, 67 slices), and protocol 5 (b-factor = 1000 s/mm$^2$, TR/TE = 8000/80.8-85.0, 67 slices). Data were acquired in either 30 or 64 directions. For the purposes of data harmonization, the 64-direction protocols were downsampled. We used a MATLAB dot product function to identify the diffusion-encoding directions that were most like those in the 30-direction protocol \cite{CBM21}. Whole-brain, resting state (rs-fMRI) blood oxygenation level-dependent (BOLD) images were collected using T2$\ast$-weighted echo-planar imaging (TR = 2000ms, voxel size = 3.75mm $\times$ 3.75mm $\times$ 3.8mm, FOV = 240mm$^2$, 64-by-64 matrix, interleaved data acquisition). Participants were asked to fixate on a crosshair with their eyes open. Additional parameters differed slightly between the first protocol (TE = 27ms, 77$^{\circ}$ flip angle, 39 slices) and the other four protocols (TE = 25ms, 90$^{\circ}$ flip angle, 35 slices). For data harmonization, the first 150 volumes were included from each protocol.

\subsection{MRI data Processing}\label{sec:MRI-processing}

\subsubsection{Diffusion data processing}

Denoising was performed using \texttt{dwidenoise} \cite{VNC16}. For every denoised B0, a ``synthetic" undistorted B0 was created using \texttt{Synb0-DisCo} \cite{SBH19}. We then used tools from FSL \cite{JBB12} and MRtrix3 \cite{TSR19} to process the denoised diffusion data as follows. FSL's topup \cite{ASA03} estimated and corrected the susceptibility-induced off-resonance field using the original and synthetic B0 images and applied the correction to the full diffusion data. The data were then run through FSL's eddy \cite{AS16}, MRtrix's dwi2mask \cite{DRC16}, and FSL's BET \cite{Sm02} (to create a combined mask), and MRtrix's dwibiascorrect (for B1 field inhomogeneity correction). Diffusion images were registered to the MNI-space standard brain using a rigid-body registration from the distortion-corrected B0 to the same subject's T1, and a non-linear registration using ANTS \cite{ATW11} \cite{WDS11} from the T1 to the FSL standard brain. 

We ran deterministic tractography using MRtrix3's tckgen, using an anatomically-constrained (ACT) procedure \cite{STC12} that uses the Freesurfer-based five-tissue-type (5TT) segmentation of the T1 image from 5ttgen \cite{STC12}. 100,000 tracks were generated from random points in the mask image at a step size of 0.2mm, with the following tckgen parameters: 60-degree maximum angle between steps and excluding tracks with lengths outside the range of 4mm-200mm. The Brainnetome atlas, containing 246 regions (210 and 36 subcortical regions) was used to define nodes \cite{FLZ16}. Streamline count between every pair of nodes was extracted into weighted, undirected connectomes using \texttt{tck2connectome} \cite{TSR19} to create a 246-by-246 matrix for each subject. We excluded edges consisting of fewer than three streamlines. Furthermore, to eliminate the effect of region size, the number of streamlines was normalized by the mean of the sizes of the two regions for each edge. 

\subsubsection{Functional Data Processing}\label{sec:functional-data-processing}

We excluded the first six volumes of the rs-fMRI data to ensure that steady-state had been reached. The processing pipeline was run using FSL v5.0.9 and included the following steps: slice-timing correction, motion correction using rigid-body transformation, intensity normalization (scaling median intensity of each subject's voxel data within the brain mask to 10000), nuisance signal regression of mean white matter and cerebral spinal fluid \cite{SJW04}, high-pass temporal filtering using a Gaussian filter with 0.01Hz cutoff, and spatial smoothing with a 6mm full width at half maximum Gaussian kernel. We then ran \texttt{ICA-AROMA} \cite{PMvR15} to remove additional noise confounds.   rs-fMRI data were registered to the T1w data and MNI template in the same manner as the diffusion data using ANTs, but with the rs-FMRI/T1w transform allowed to be a 12-degree-of-freedom affine transform. All participants had a mean relative motion of $< 0.4 $mm, measured using FSL's MCFLIRT using the middle volume as the reference. Using the Brainnetome atlas \cite{fan2016human}, we obtained the timecourse of brain activity from each region across 144 time points and computed correlations between all possible pairs of regions. A Fisher transform was performed on correlation matrices. Correlations between nodes less than 20mm apart were not calculated, as close physical proximity may artificially inflate the correlation values \citep{power2012spurious}. These procedures resulted in a $246\times246$ correlation matrix.

\subsection{Overview of Tensor-based Methods}\label{sec:tensor-formatted-data}

Our data consisted of both functional MRI and diffusion MRI data for 275 individuals. For either modality of MRI data of every individual, the data was a symmetric matrix showing connectivity (either streamline count or functional correlation) between each of the 246 regions. Therefore, for both functional and diffusion data, there were $275 \times 246 \times 246$ values. We centralized and scaled the values to zero mean unit variance for both functional and diffusion connectivity matrices so that the scale of these values became similar.

\begin{figure}[!htb]
	\begin{center}
		\includegraphics[width=.48\linewidth]{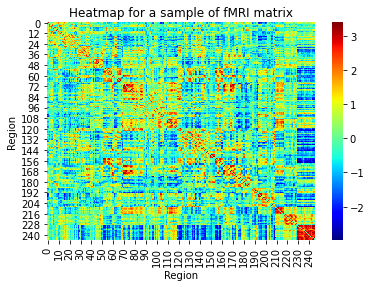}
		\includegraphics[width=.48\linewidth]{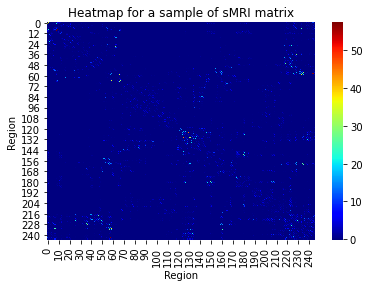}
		\caption{Heatmaps of one participant's functional connectivity matrix (left) and diffusion connectivity matrix (right). A darker colored cell means a stronger connection between the corresponding ROIs.}
		\label{mridata}
	\end{center}
\end{figure}

Figure \ref{mridata} provides the heatmaps of the functional and diffusion matrix of one subject to illustrate the difference between matrices. From the heatmap of the fMRI matrix, we can see a hidden block structure inside the matrix. In the middle of the matrix, many regions have strong positive correlations. For the dMRI matrix, however, we do not have such an apparent structure. The matrix itself is sparse, as is expected in the human brain with estimates of 30\% connection density for white matter \citep{roberts2017consistency,buchanan2020effect}.


\subsection{Description of Tensor-based Methods}\label{sec:method}

\begin{figure}[h]
	\begin{center}
		\includegraphics[scale=0.45]{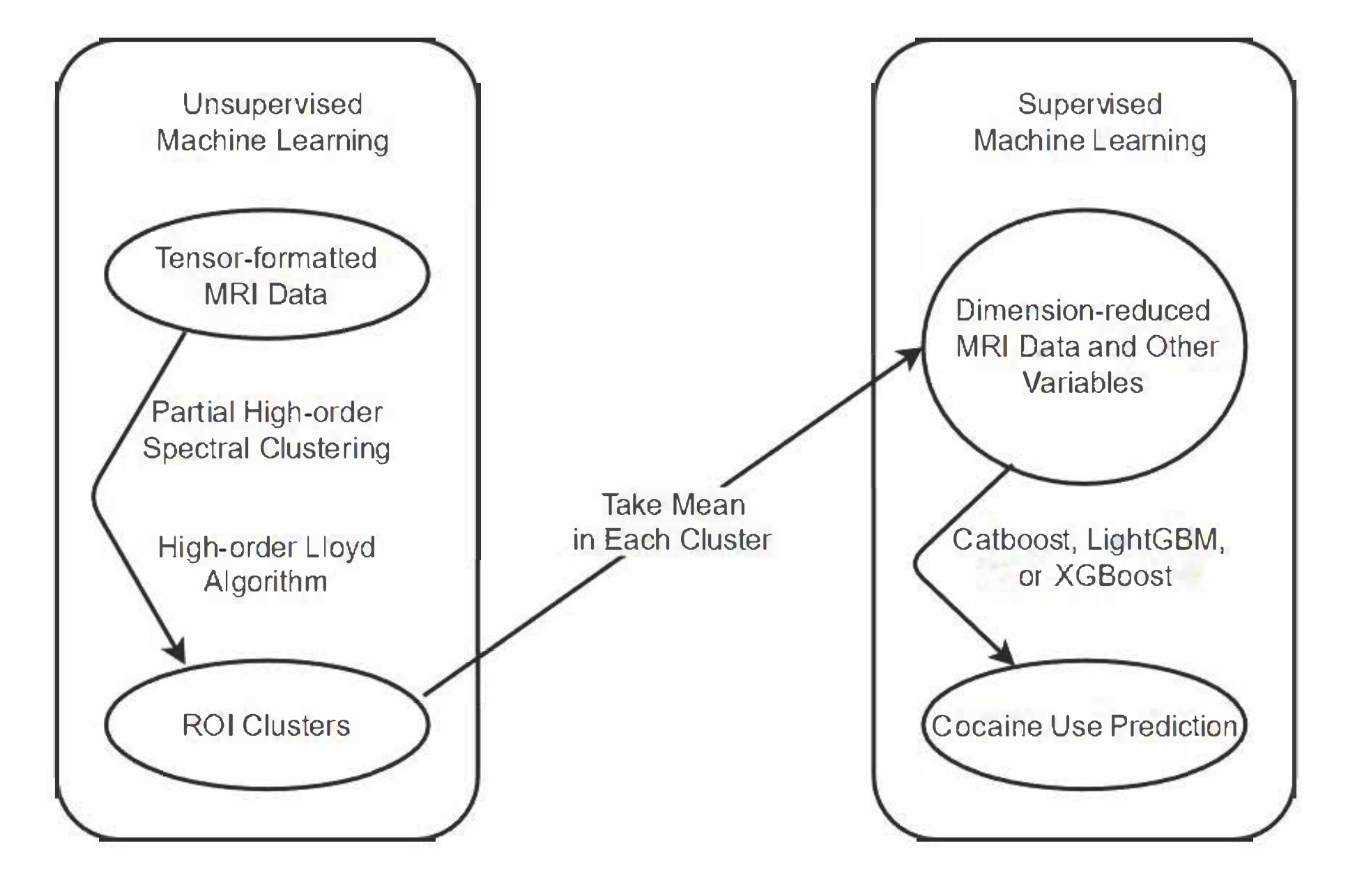}
		\caption{Flowchart of the proposed tensor-based method}
		\label{outline}
	\end{center}
\end{figure}

\begin{figure}
	\begin{center}
		\subfigure[Step 1. Acquisition of functional and diffusion MRI connectomes]{\includegraphics[height=.7in]{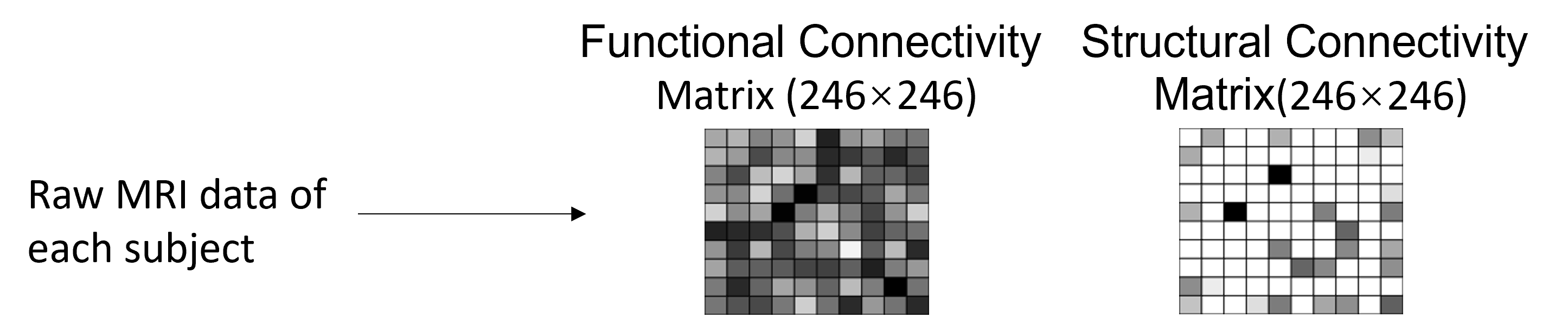}}\\
		\subfigure[Step 2. Tensor formalization]{\includegraphics[height=1.7in]{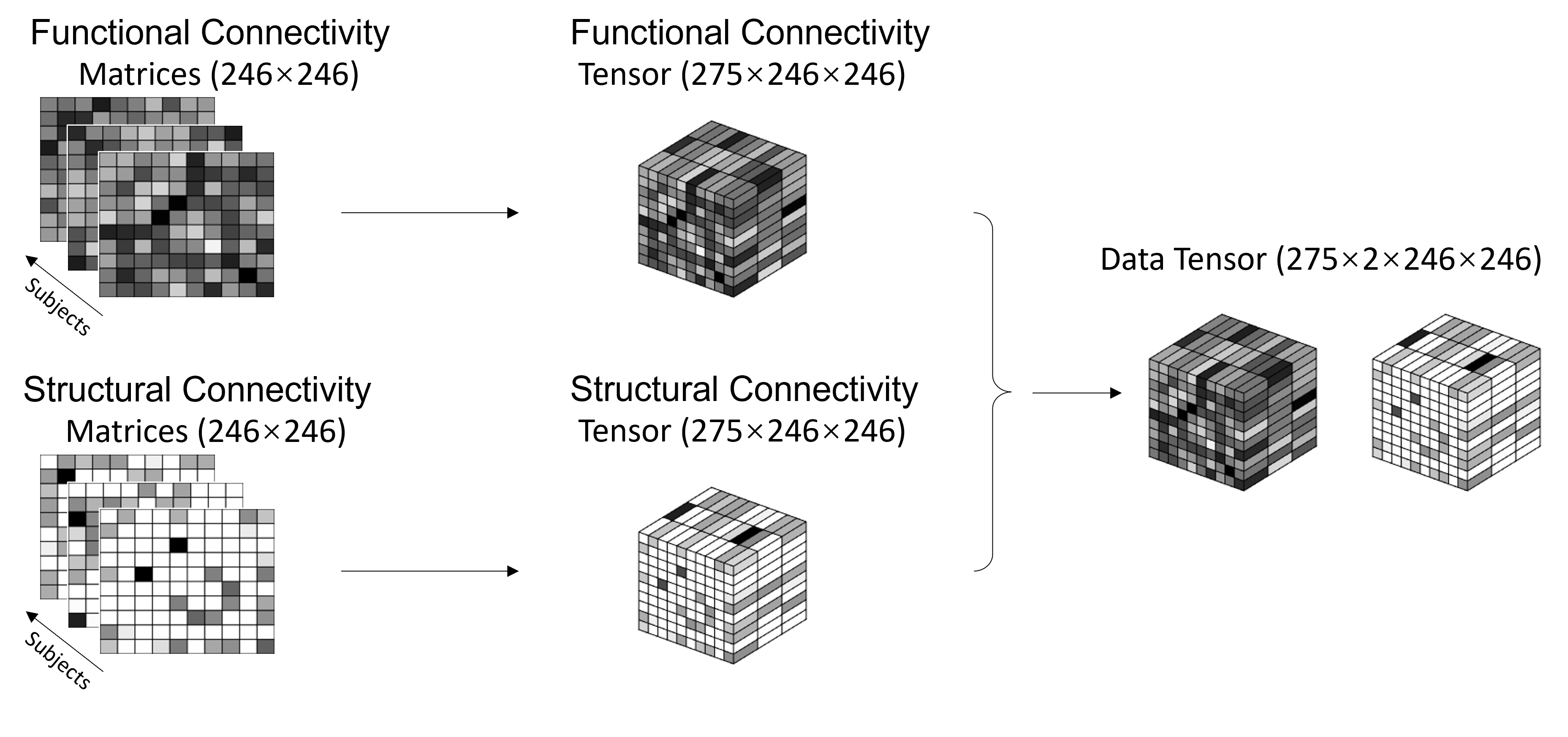}}\\
		\subfigure[Step 3. Apply tensor-based machine learning on data tensor to cluster ROIs]{\includegraphics[height=1.4in]{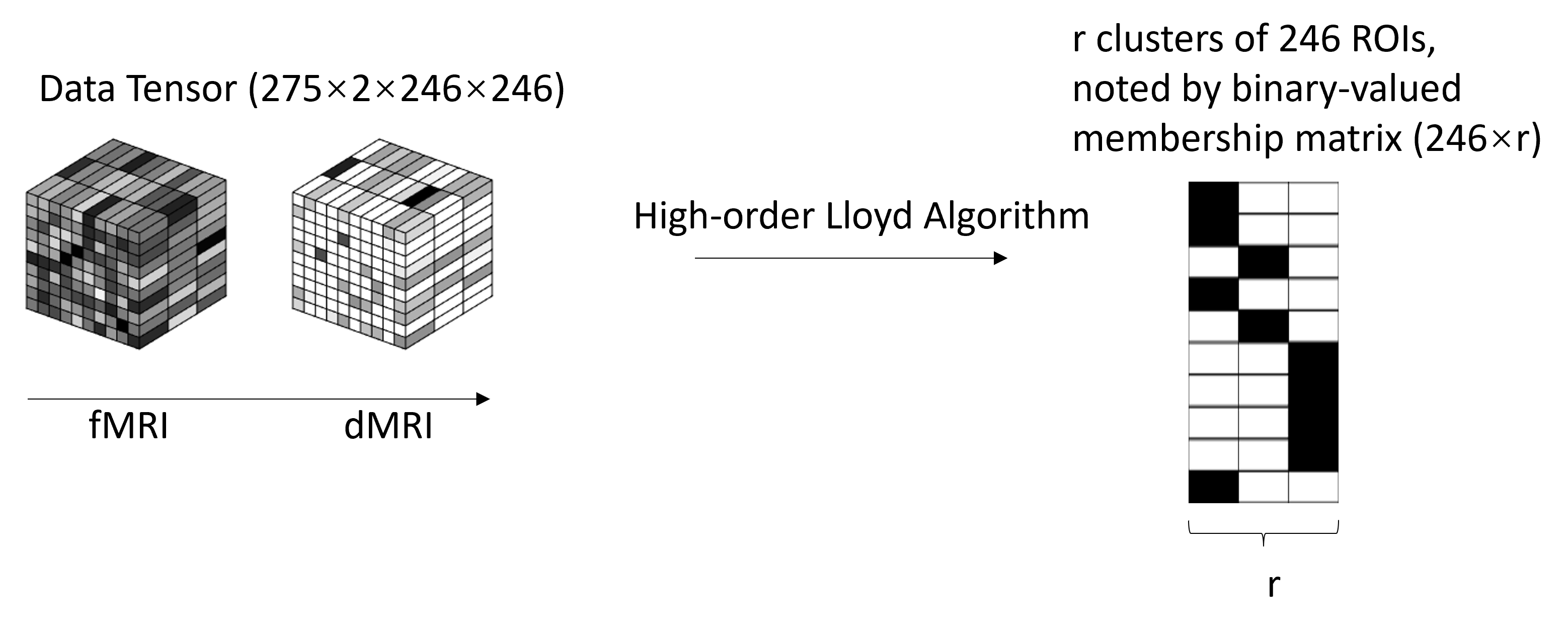}}\\
		\subfigure[Step 4. Dimension reduction based on ROI clusters]{~\quad \includegraphics[height=1.5in]{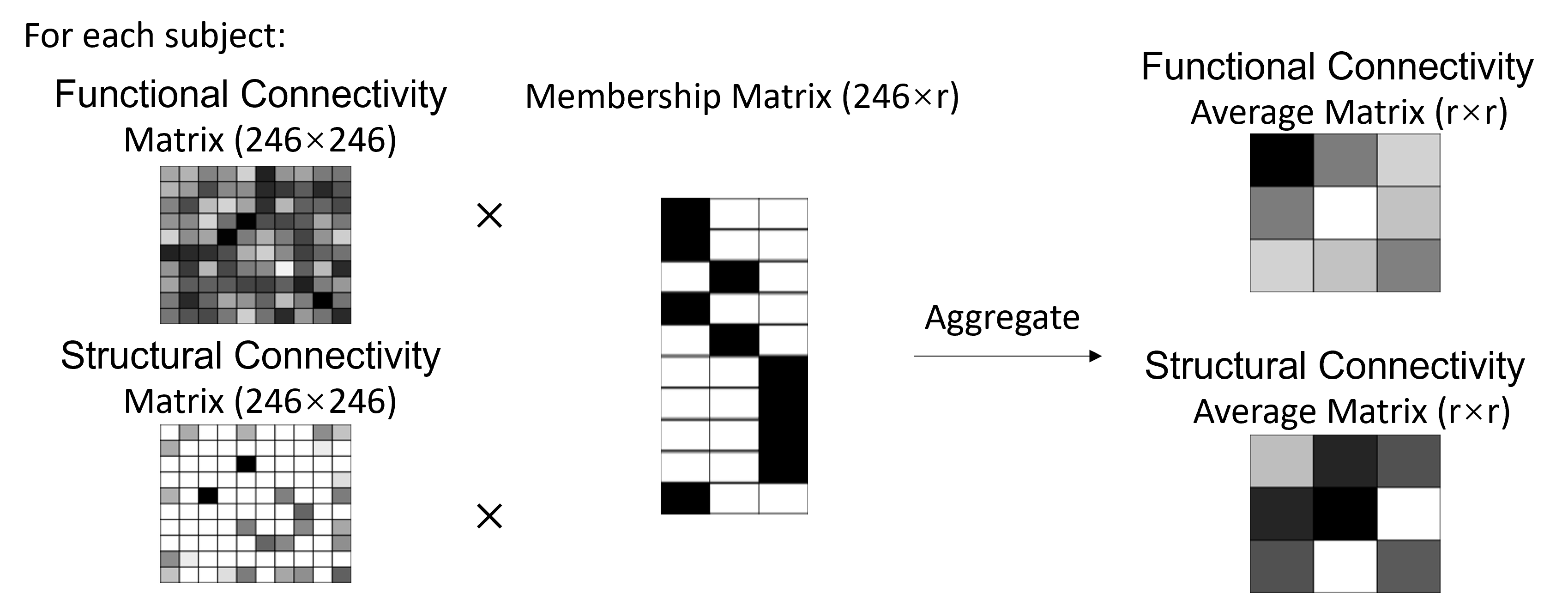}\quad~}\\	
		\subfigure[Steps 5. Apply supervised learning on dimension reduced connectome data]{~\quad \includegraphics[height=1in]{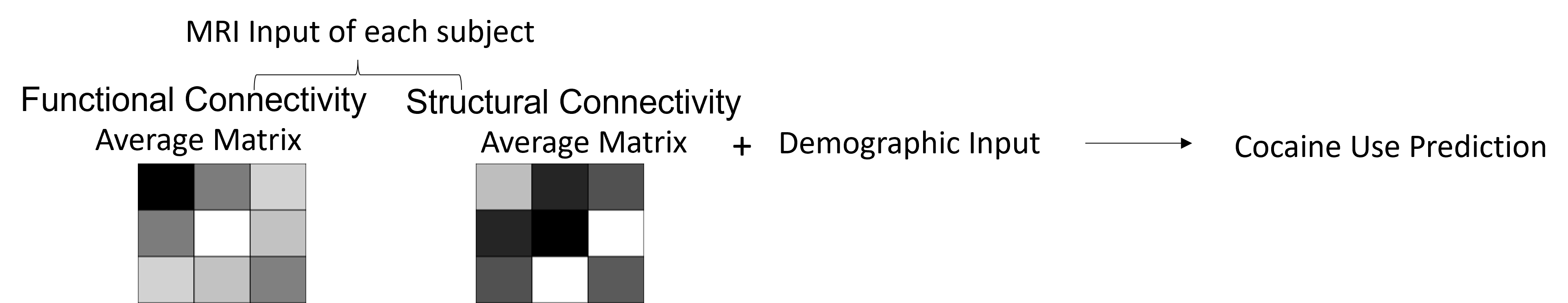}\quad~}	
	\end{center}
	\caption{Illustration of Methods}
	\label{fig:procedure}
\end{figure}

Our method includes two steps: first, an unsupervised machine learning algorithm was applied to reduce the dimension of the data; second, a supervised machine learning algorithm was applied to achieve the prediction of cocaine use. Figure \ref{outline} gives a flowchart and Figure \ref{fig:procedure} illustrates each step of our procedure.

\subsubsection{Unsupervised Learning on Connectivity Tensors}\label{sec:uml}

We applied an unsupervised machine learning procedure to extract key and interpretable information from the data and reduce its dimension. The procedure is illustrated in Figure \ref{fig:procedure} (a)-(d).

Denote $\cY \in \mathbb{R}^{275 \times 2 \times 246 \times 246}$ the connectome tensor, where the four directions of the tensor $\cY$ correspond to different subjects (275), different modalities (functional or diffusion MRI), different ROIs (246), and different ROIs (246) again, respectively. For example, let $k$ refer to the functional modality, then the value $\cY_{n,k,i,j}$ refers to the $(i,j)$th entry of subject $n$'s functional connectivity matrix. We aim to obtain valid and interpretable clustering of ROIs in this step. 
It is known that brain regions form modules: different ROIs belonging to the same cluster have similar connectivity patterns. This implies, relative to the brain connectome data, the ROIs belonging to the same cluster have similar average connectome values. Suppose the 246 ROIs in the human brain approximately form $r$ clusters denoted by a vector $z\in \{1,\dots,r\}^{246}$: the $j$-th entry of $z$ equals to $m$ if and only if the $j$-th entity belongs to the $m$-th cluster. Then we have
$$\cY_{i_1, i_2, i_3, i_4} = \cS_{i_1, i_2, z_{i_3}, z_{i_4}} + \cE_{i_1,i_2,i_3,i_4},$$
where $\cS \in \mathbb{R}^{275\times 2\times r\times r}$ denotes the average connectome values between ROIs belonging to clusters $z_{i_3}$ and $z_{i_4}$ in subject $i_1$, connectome data modality $i_2$, $\cE$ is the residual of this approximation with presumably smaller amplitude. It is worth mentioning that after proper permutation of ROI indices, the connectome tensor $\cY$ would show blockwise patterns \cite{Han20}. Utilizing the tensor notation and algebra (see details in Section \ref{sec:algorithms}), the connectome tensor $\cY$ shows a blockwise structure as follows:
\[ \cY=\cS \times_{3} M \times_{4} M + \mathcal{E},\]
where $\cS \in \mathbb{R}^{275 \times 2 \times r \times r}$ is a core tensor mentioned above and $M \in \{0,1\}^{246 \times r}$ is the membership matrix. In particular, in $j$-th row of $M$, there is one value of 1 at column $z_j$ and all the other values are 0. There is a one-to-one correspondence between matrices $M$ and the membership vector $z$: $M_{j, m} = 1$ if and only if $(z)_j=m$. 



Then we applied the tensor-based  clustering algorithm to obtain the clusters $z$. Our algorithm consists of two parts. In the first part, we initialized the clusters using the partial high-order spectral clustering  \cite{Han20}, which are based on two key ingredients: high-order SVD \cite{DLL00} and $k$-means clustering \cite{lloyd1982least}. Second, we applied the high-order Lloyd algorithm \cite{Han20} to iteratively refine our initialization and obtain the final clusters. The pseudocode of high-order SVD and high-order Lloyd are summarized in Algorithms \ref{alg:hlloyd} and \ref{alg:phsc} in Section \ref{sec:algorithms} in the Appendix. 

We chose $r=$ 5--14 based on the Bayesian information criterion (BIC) discussed in \cite{Han20}.
Our output of initialization step were memberships $z^{(0)} \in  \{1,\dots,r\}^{246}$. With $z^{(0)}$, we were ready to apply Algorithm \ref{alg:hlloyd} to obtain the final clustering membership labels $z$. 

\subsubsection{Supervised Machine Learning on Connectivity Tensors}\label{sec:method-supervised-learning}

The goal of the supervised machine learning procedure is to make a prediction for cocaine status (labeled as 1 for cocaine users and 0 otherwise). The clustering result in the unsupervised machine learning rendered $r = 5,\ldots, 12$ clusters of ROIs. For each individual and for each fMRI/dMRI matrix, there are $r \times r$ blocks. We took the mean of each block and obtain a $r \times r$ symmetric matrix. Removing the duplicate values in this matrix, we had $r(r+1)/2$ numbers. For each individual, we combined the $r(r+1)/2$ values from fMRI data and another $r(r+1)/2$ values from dMRI data. Therefore, our features included $r(r+1)$ values from fMRI and dMRI data, and participant characteristics (age, gender, race, HIV status). Our input data had 275 observations and $r(r+1)+4$ features. See an illustration of the procedure in Figure \ref{fig:procedure}(e). 

We trained the aforementioned machine learning models, \texttt{Catboost} \cite{PGV17}, \texttt{LightGBM} \cite{Ke17} and \texttt{XGBoost} \cite{CG16} with 10-fold nested cross-validation on the dataset. These algorithms are for gradient boosting on decision trees. Gradient boosting is one of the most powerful machine learning algorithms to date, and the three models are among the most popular algorithms in data science.  


To train our binary classifier, we used the method of Nested cross-validation to avoid overfitting. We divided the sample equally into 10 folds and trained the model 10 times. At each time, we held 1 fold as the test set and the other 9 folds as the training set. We trained the models by boosting (implemented by \texttt{Catboost}, \texttt{LightGBM}, or \texttt{XGBoost} in Python) on the training set. We divided the training set into 3 subfolds to tune hyper-parameters by cross-validation (2 used as a sub-training set and 1 used as a validation set). For each choice of hyper-parameters and training set, we fit the model with the sub-training set and got a prediction accuracy score on the validation set. We took the average of three accuracy scores (by rotating 3 subfolds as a validation set) for each choice of hyper-parameters. Our final model corresponded to the choice of hyper-parameters that achieve the highest average accuracy score.

Once our model was fixed, we returned to the whole training set (consisting of 9 folds) and test set. We fit our model with the training set and get a prediction accuracy score on the test set. We repeated the procedure by rotating 10 folds as a test set and then averaged these 10 prediction accuracy scores to determine the accuracy scores for this cross-validation. We repeated the above cross-validation 100 times and recorded the mean, maximum, and minimum of all 100 prediction accuracy scores.

\section{Results}\label{sec:results}

\subsection{Unsupervised Learning}\label{sec:result-unsupervised}

Participant characteristics are listed in Table \ref{descriptivetable}. The sample included 275 participants, of whom 83 (42\%) were categorized as cocaine users. Participants were predominantly male (71\%), African-American (74\%), and HIV-positive (63\%) with a mean age of 41.16 (SD = 9.07). Among cocaine users, participants had used on an average of 31.80 days (SD = 25.52) of the past 90 and had used regularly for 16.11 years (SD = 8.67), and 76 (92\%) endorsed diagnostic criteria for current cocaine use dependence. On the day of the MRI, 64 (83\%) had a cocaine-positive urine drug test.  

\begin{table}[!ht]
	\centering
	\begin{tabular}{l|ll}\toprule
		&\makecell[l]{Cocaine users\\ (N = 83)} & \makecell[l]{Non-cocaine \\ users (N = 192)}  \\ \midrule
		Age, M (SD) & 45.83 (7.29) & 39.15 (9.01)  \\ 
		HIV positive, \%, N & 57.83\%, 48 & 65.10\%, 125\\
		\makecell[l]{Male Gender, \%, N} & 68.67\%, 57 & 71.35\%, 137  \\ 
		Race & ~ & \\ 
		\makecell[l]{$\qquad$Black/African American,\\ $\qquad$\%, N} & 87.95\%, 73 & 67.71\%, 130 \\ 
		\makecell[l]{$\qquad$Other/Mixed,\\ $\qquad$\%, N} & 12.20\%, 10 & 32.29\%, 62 \\ \bottomrule
	\end{tabular}
	
	\caption{Descriptive Statistics} \label{descriptivetable}
\end{table}

\begin{figure}[ht!]
	\minipage{0.44\textwidth}
	\includegraphics[width=\linewidth]{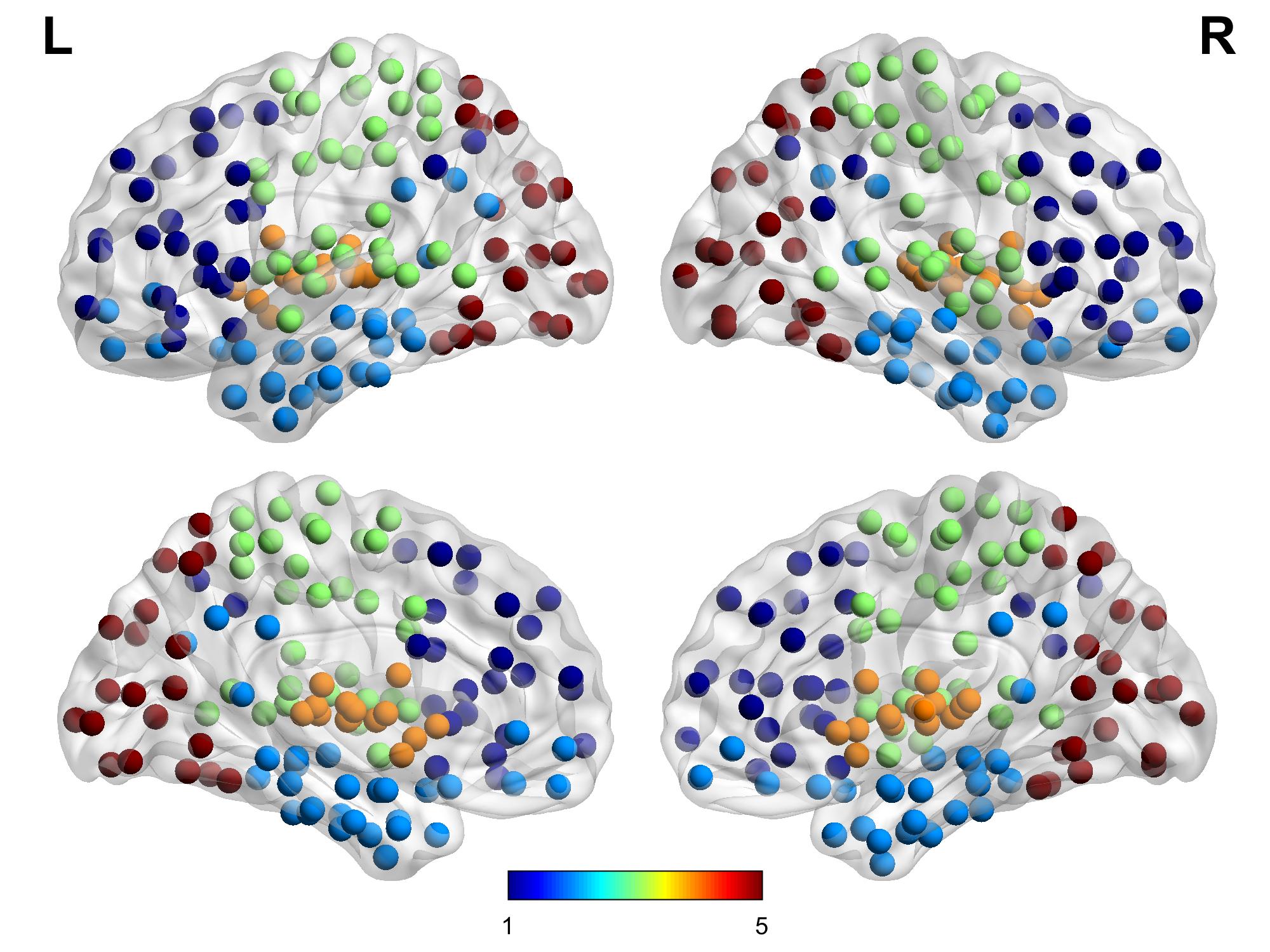}
	\endminipage \hspace{1mm}  \hfill
	\minipage{0.44\textwidth}
	\includegraphics[width=\linewidth]{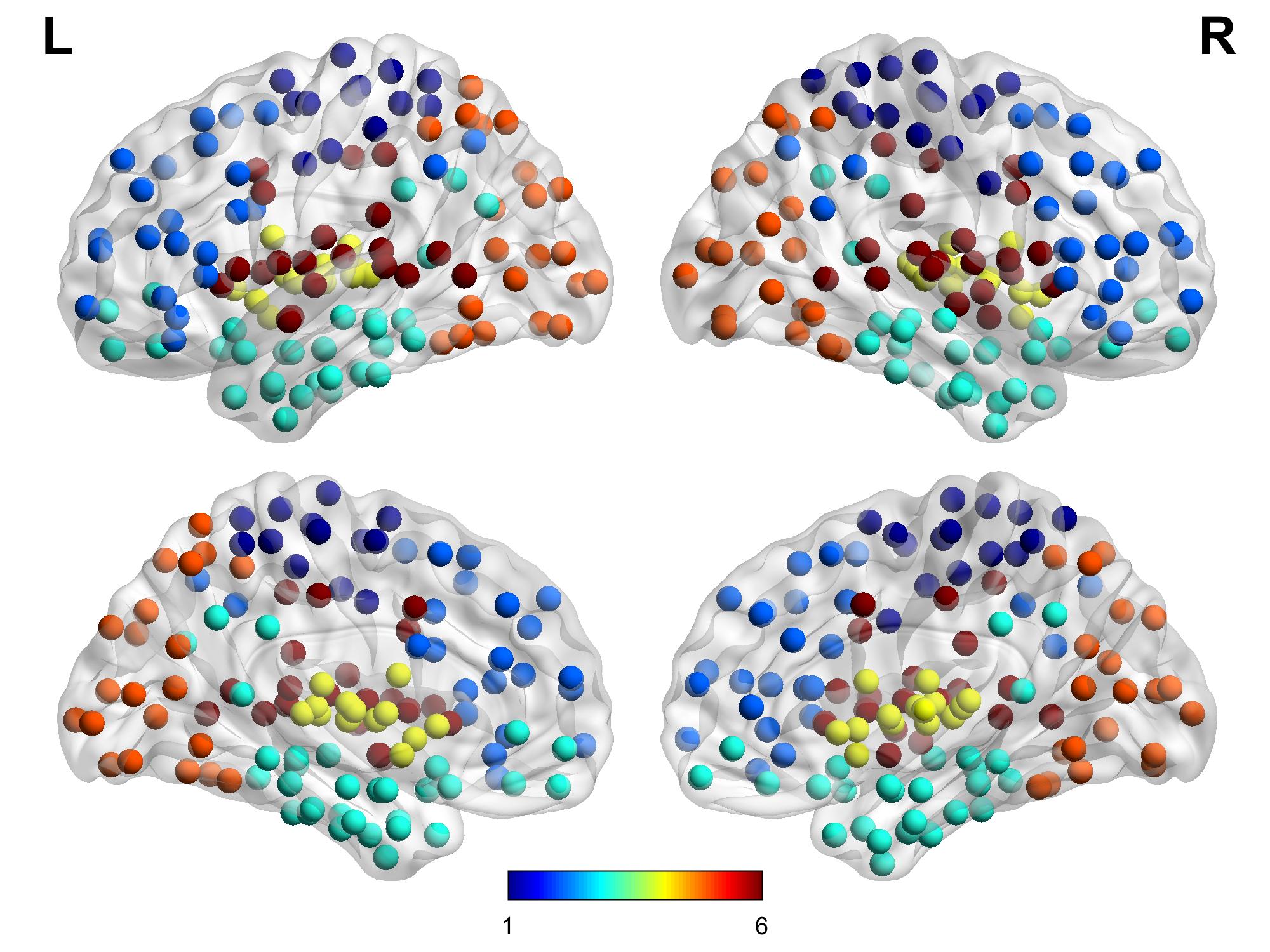}
	\endminipage \\
	\minipage{0.44\textwidth}
	\includegraphics[width=\linewidth]{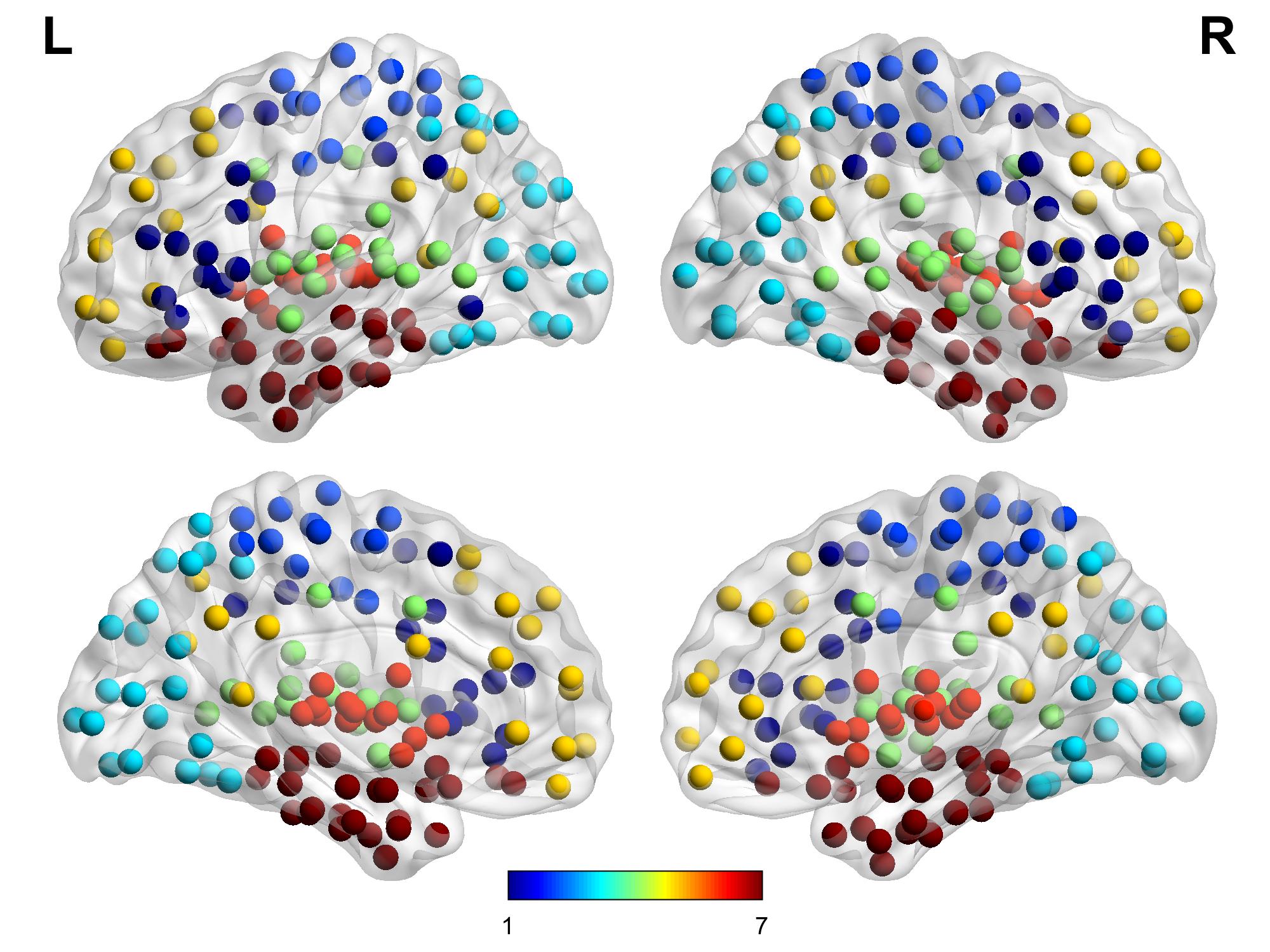}
	\endminipage \hspace{1mm}  \hfill
	\minipage{0.44\textwidth}
	\includegraphics[width=\linewidth]{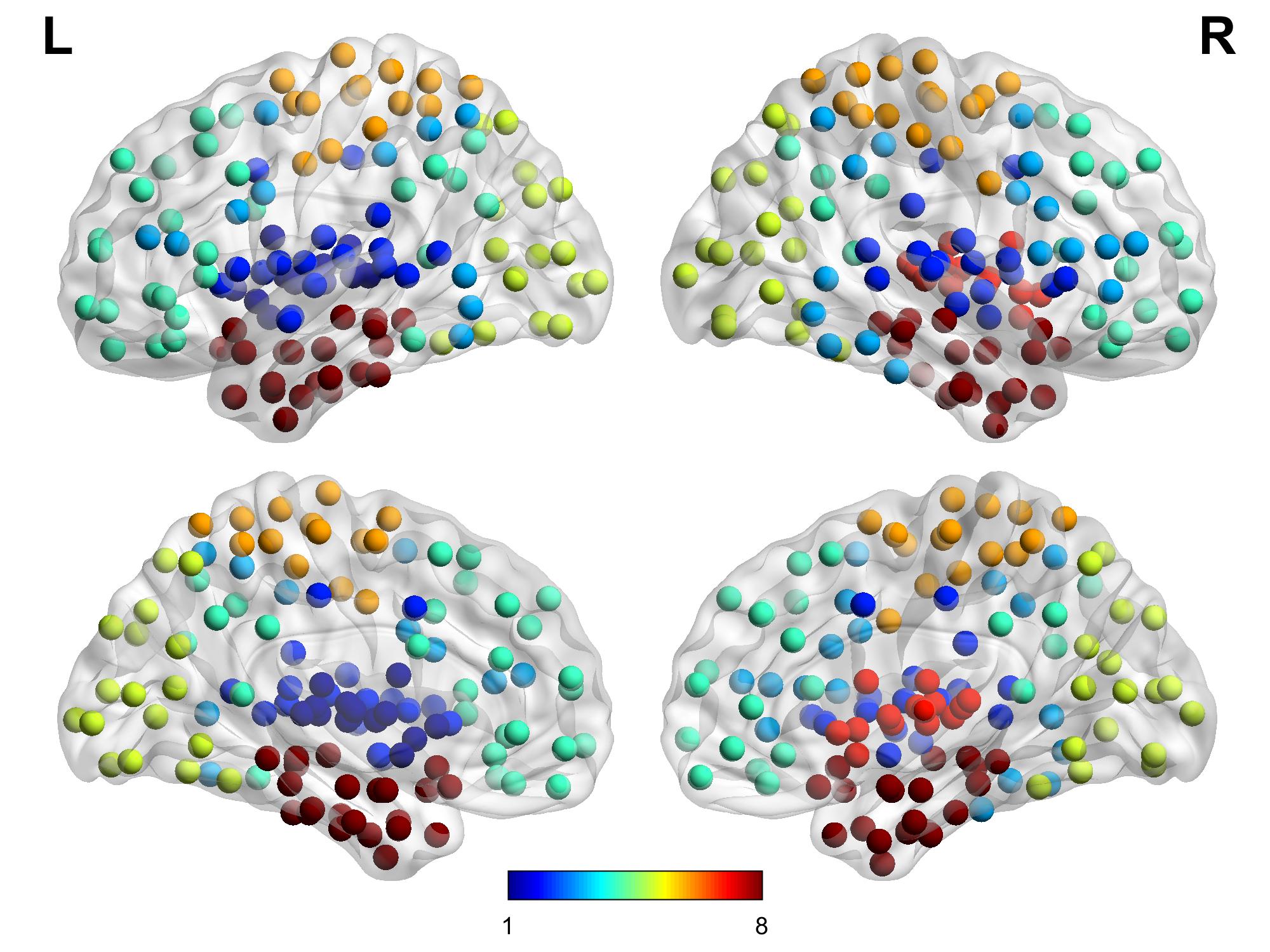}
	\endminipage \\
	\minipage{0.44\textwidth}
	\includegraphics[width=\linewidth]{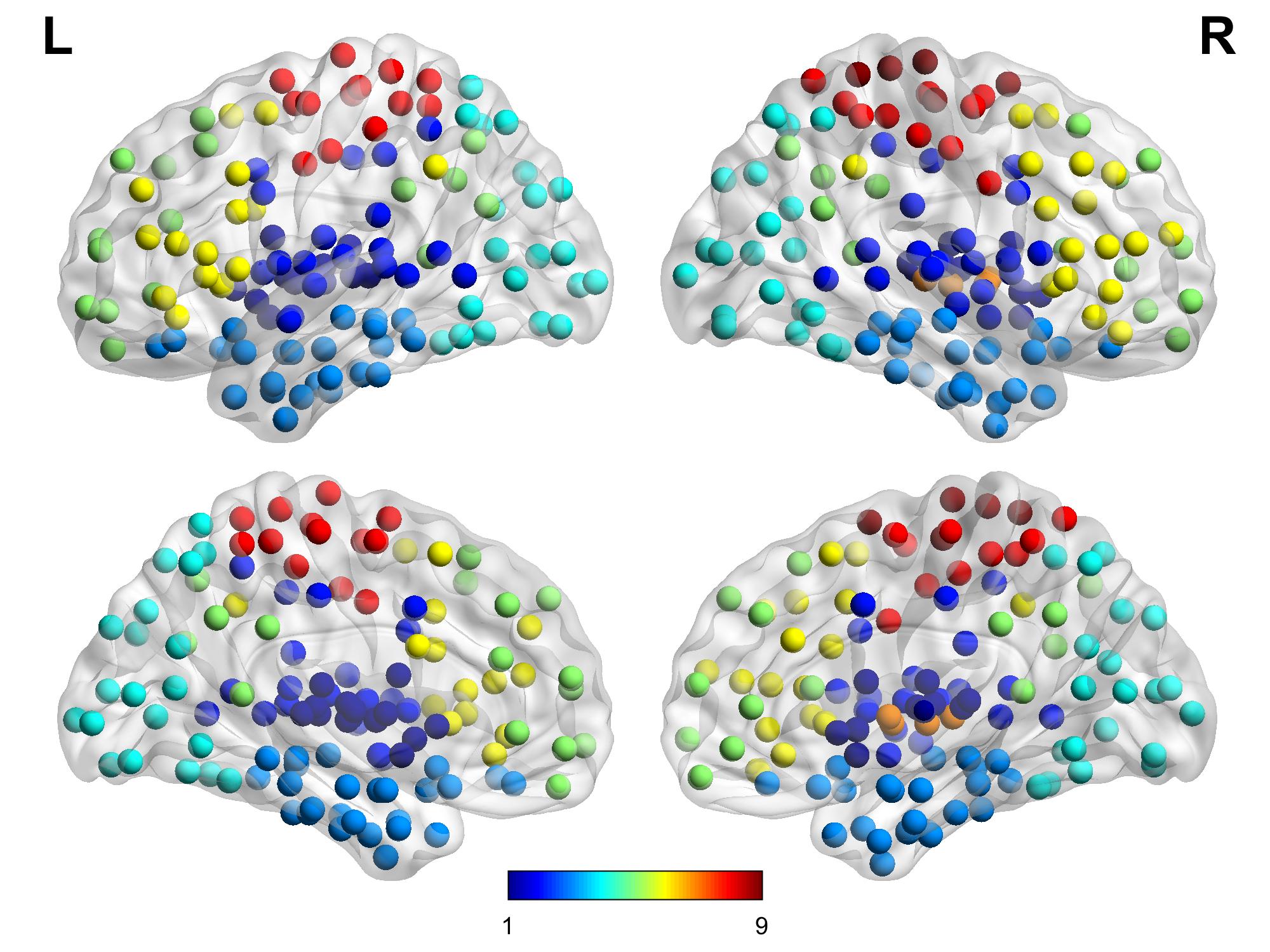}
	\endminipage \hspace{1mm}  \hfill
	\minipage{0.44\textwidth}
	\includegraphics[width=\linewidth]{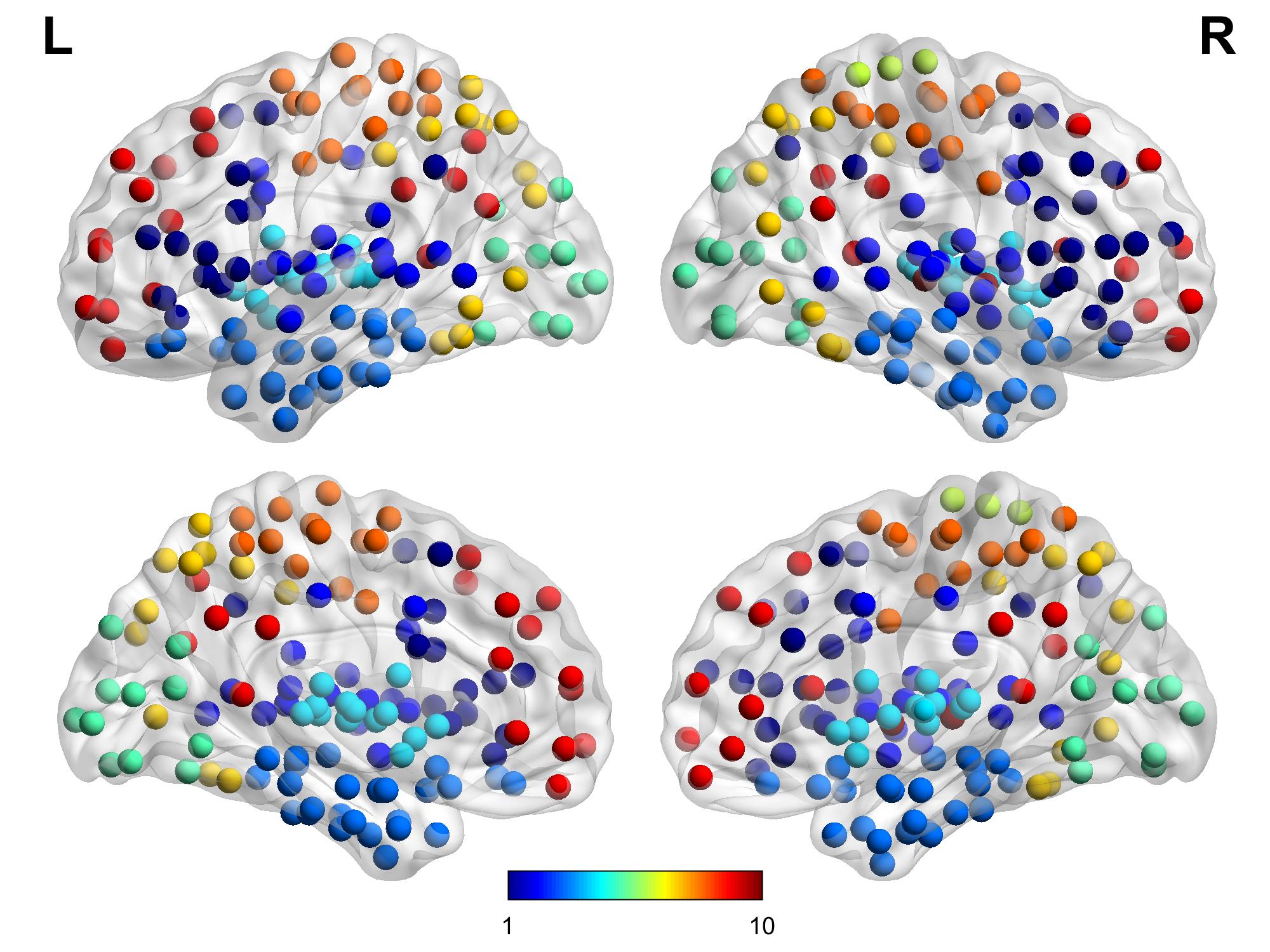}
	\endminipage \\
	\minipage{0.44\textwidth}
	\includegraphics[width=\linewidth]{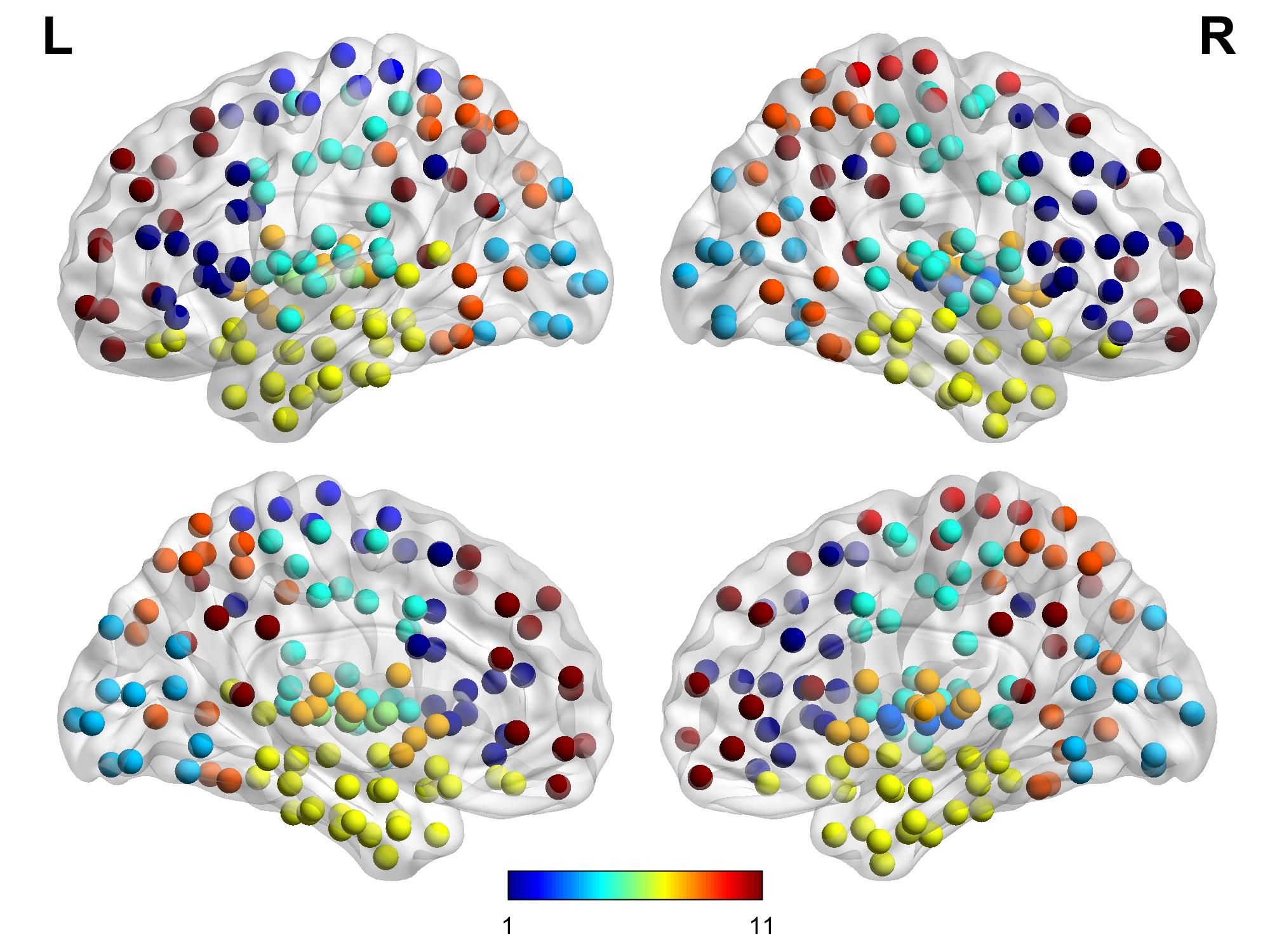}
	\endminipage \hspace{1mm}  \hfill
	\minipage{0.44\textwidth}
	\includegraphics[width=\linewidth]{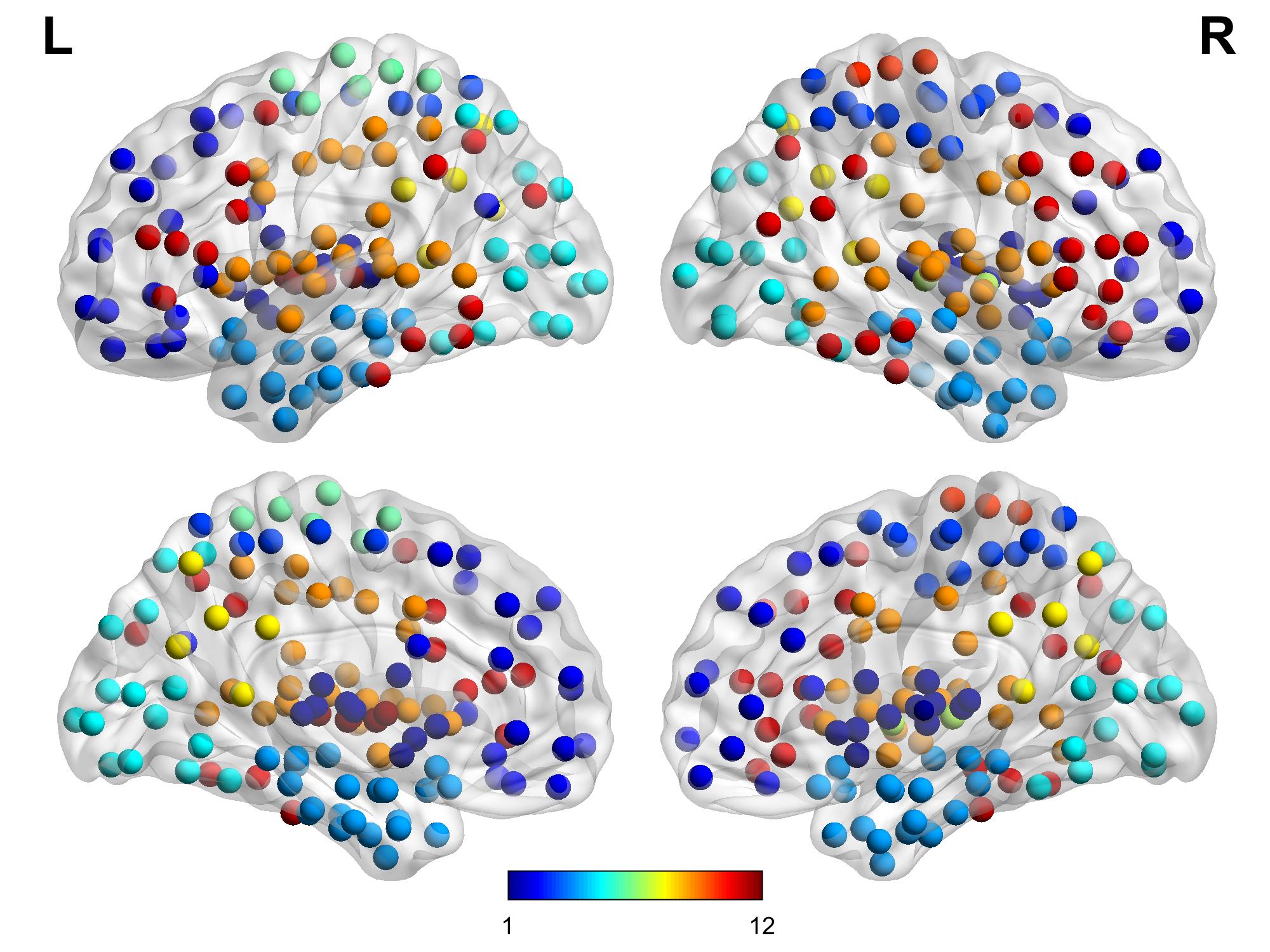}
	\endminipage 
	\caption{Clustering visualization for the number of clusters 5-12}
	\label{clusterfigure}
\end{figure}

\begin{figure}[ht!]
	\minipage{0.44\textwidth}
	\includegraphics[width=\linewidth]{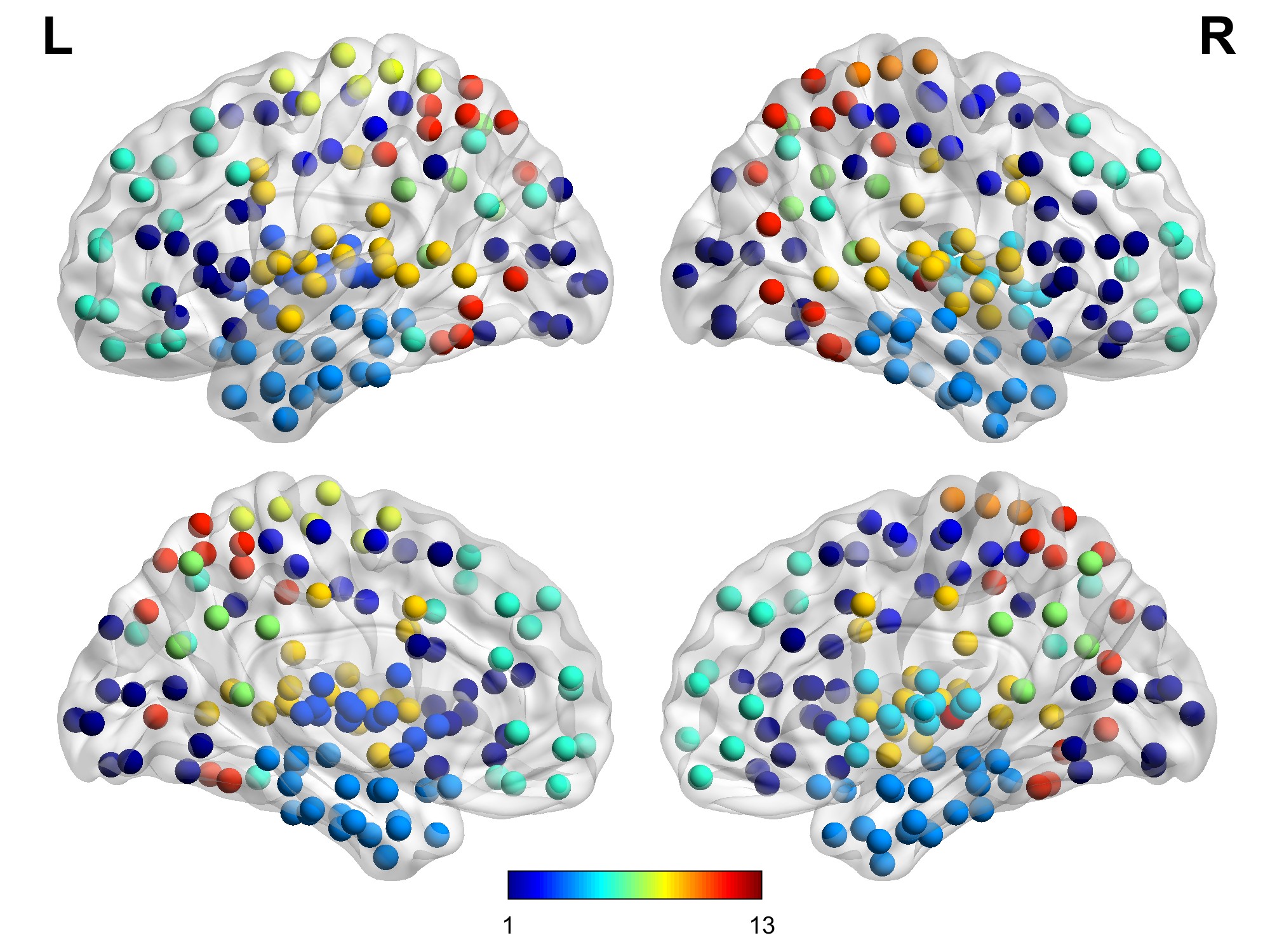}
	\endminipage \hspace{1mm}  \hfill
	\minipage{0.44\textwidth}
	\includegraphics[width=\linewidth]{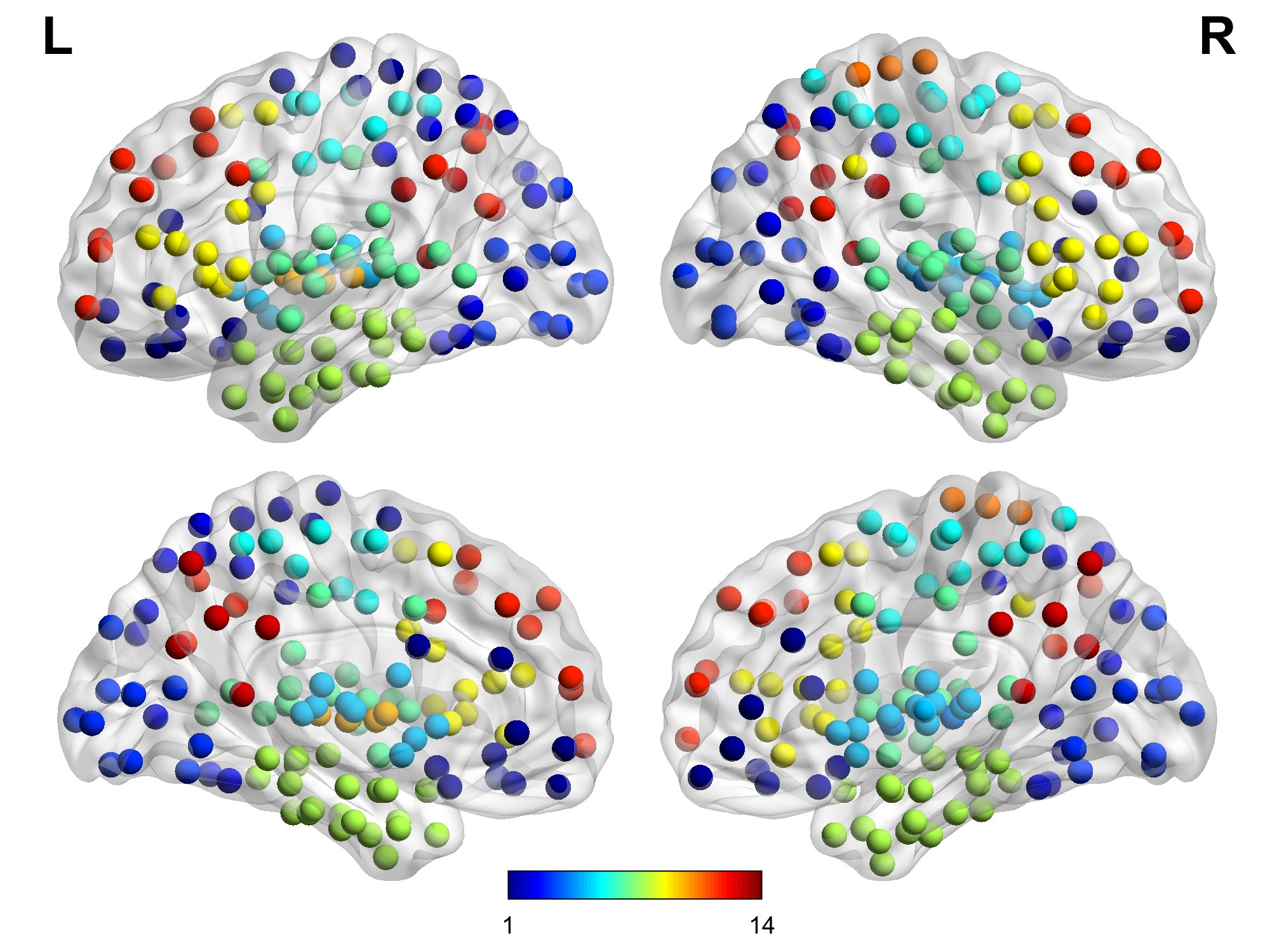}
	\endminipage
	\caption{Clustering visualization for the number of clusters 13 and 14}
	\label{clusterfigure_1}
\end{figure}

\begin{figure}[!htb]
	\begin{center}
		\includegraphics[width=.6\linewidth]{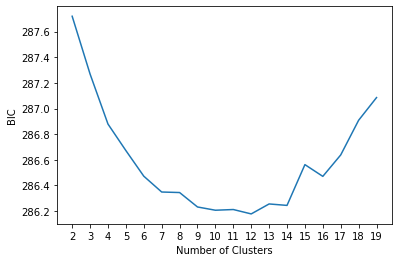}
		\caption{BIC of clustering result with different number of clusters}
		\label{fig:BIC}
	\end{center}
\end{figure}

Next, we present the clustering results for a range of choices for the number of clusters, specifically, 5 to 14. Our selection of clusters is confined within the range of 5 to 14 based on the BIC criterion \cite{Han20}. The membership of all ROIs for each configuration of the number of clusters is listed in Table \ref{table_clustering_result}. To provide a visual representation of the results, Figure \ref{clusterfigure} and Figure \ref{clusterfigure_1} display the clustering outcomes, with each color representing a distinct cluster. Figure \ref{fig:BIC} provides the BIC for different choices of the number of clusters. We pick the number of clusters with low BIC. 


\subsection{Supervised Learning}\label{sec:results-supervised}

We further apply the supervised machine learning method described in Section \ref{sec:method-supervised-learning}. For each choice of the number of clusters and the choice of the machine learning algorithm, we fit the model as in Section \ref{sec:method-supervised-learning} and obtained average value, maximum value, and minimum value of the following four metrics: cocaine group prediction accuracy, area under the ROC curve (AUROC), area under the precision-recall curve (AUPRC), and the maximum of minimum precision-recall pair on the precision-recall curve Min(Re, P+). The first two are widely used and the last is the minimum of recall that is greater than precision in the precision-recall curve. See \cite{fogarty2005case, davis2006relationship} for detailed explanation and examples of those metrics. Tables \ref{accuracytable}, \ref{auroctable}, \ref{auprctable} and \ref{minreprtable} show the results with different predictive metrics for each machine learning algorithm with clusters from 5-9. When clusters are more than 9, the metrics will not be improved and some clusters would correspond to multiple regions in the brain, which is not as interpretable as the cases with clusters from 5-9.

\begin{table}[!ht]
	\begin{center}
		\begin{tabular}{p{17.5mm}|p{21mm}|p{21mm}|p{21mm}|p{21mm}|p{21mm}}
			\hline
			\# clusters & 5 & 6 & 7 & 8 & 9 \\
			\hline
			\texttt{Catboost} & \begin{tabular}{@{}c@{}}0.847 \\ (0.866,0.829)\end{tabular} & \begin{tabular}{@{}c@{}}0.857 \\ (0.880,0.833)\end{tabular} & \begin{tabular}{@{}c@{}}0.846 \\ (0.866,0.833)\end{tabular} & \begin{tabular}{@{}c@{}}0.846 \\ (0.869,0.822)\end{tabular} & \begin{tabular}{@{}c@{}}0.851 \\ (0.873,0.825)\end{tabular}\\
			\hline
			\texttt{LightGBM}   & \begin{tabular}{@{}c@{}}0.826 \\ (0.851,0.785)\end{tabular} & \begin{tabular}{@{}c@{}}0.852 \\ (0.876,0.818)\end{tabular} & \begin{tabular}{@{}c@{}}0.853 \\ (0.876,0.832)\end{tabular} & \begin{tabular}{@{}c@{}}0.844 \\ (0.869,0.811)\end{tabular} & \begin{tabular}{@{}c@{}}0.851 \\ (0.876,0.821)\end{tabular} \\
			\hline
			\texttt{XGBoost}  & \begin{tabular}{@{}c@{}}0.831 \\ (0.851,0.808)\end{tabular} & \begin{tabular}{@{}c@{}}0.849 \\ (0.884,0.825)\end{tabular} & \begin{tabular}{@{}c@{}}0.851 \\ (0.876,0.829)\end{tabular} & \begin{tabular}{@{}c@{}}0.853 \\ (0.887,0.829)\end{tabular} & \begin{tabular}{@{}c@{}}0.850 \\ (0.873,0.832)\end{tabular}   \\
			\hline
		\end{tabular}
		\caption{Mean of accuracy (with maximum accuracy and minimum accuracy in the bracket) for each number of clusters and each machine learning algorithm}
		\label{accuracytable}
	\end{center}
\end{table}

\begin{table}[!ht]
	\begin{center}
		\begin{tabular}{p{17.5mm}|p{21mm}|p{21mm}|p{21mm}|p{21mm}|p{21mm}}
			\hline
			\# clusters & 5 & 6 & 7 & 8 & 9 \\
			\hline
			\texttt{Catboost} & \begin{tabular}{@{}c@{}}0.776 \\ (0.810,0.739)\end{tabular} & \begin{tabular}{@{}c@{}}0.788 \\ (0.821,0.747)\end{tabular} & \begin{tabular}{@{}c@{}}0.768 \\ (0.801,0.739)\end{tabular} & \begin{tabular}{@{}c@{}}0.773 \\ (0.816,0.732)\end{tabular} & \begin{tabular}{@{}c@{}}0.775 \\ (0.805,0.741)\end{tabular}\\
			\hline
			\texttt{LightGBM}   & \begin{tabular}{@{}c@{}}0.762 \\ (0.803,0.711)\end{tabular} & \begin{tabular}{@{}c@{}}0.796 \\ (0.834,0.759)\end{tabular} & \begin{tabular}{@{}c@{}}0.787 \\ (0.836,0.754)\end{tabular} & \begin{tabular}{@{}c@{}}0.780 \\ (0.829,0.733)\end{tabular} & \begin{tabular}{@{}c@{}}0.786 \\ (0.823,0.755)\end{tabular} \\
			\hline
			\texttt{XGBoost}  & \begin{tabular}{@{}c@{}}0.768 \\ (0.795,0.739)\end{tabular} & \begin{tabular}{@{}c@{}}0.788 \\ (0.843,0.749)\end{tabular} & \begin{tabular}{@{}c@{}}0.782 \\ (0.822,0.746)\end{tabular} & \begin{tabular}{@{}c@{}}0.791 \\ (0.830,0.749)\end{tabular} & \begin{tabular}{@{}c@{}}0.783 \\ (0.826,0.753)\end{tabular}   \\
			\hline
		\end{tabular}
		\caption{Mean of AUROC (with maximum AUROC and minimum AUROC in the bracket) for each number of clusters and each machine learning algorithm}
		\label{auroctable}
	\end{center}
\end{table}

\begin{table}[!ht]
	\begin{center}
		\begin{tabular}{p{17.5mm}|p{21mm}|p{21mm}|p{21mm}|p{21mm}|p{21mm}}
			\hline
			\# clusters & 5 & 6 & 7 & 8 & 9 \\
			\hline
			\texttt{Catboost} & \begin{tabular}{@{}c@{}}0.786 \\ (0.833,0.724)\end{tabular} & \begin{tabular}{@{}c@{}}0.806 \\ (0.854,0.742)\end{tabular} & \begin{tabular}{@{}c@{}}0.789 \\ (0.840,0.738)\end{tabular} & \begin{tabular}{@{}c@{}}0.786 \\ (0.836,0.716)\end{tabular} & \begin{tabular}{@{}c@{}}0.799 \\ (0.843,0.742)\end{tabular}\\
			\hline
			\texttt{LightGBM}   & \begin{tabular}{@{}c@{}}0.750 \\ (0.808,0.670)\end{tabular} & \begin{tabular}{@{}c@{}}0.792 \\ (0.839,0.745)\end{tabular} & \begin{tabular}{@{}c@{}}0.795 \\ (0.844,0.745)\end{tabular} & \begin{tabular}{@{}c@{}}0.780 \\ (0.829,0.720)\end{tabular} & \begin{tabular}{@{}c@{}}0.792 \\ (0.843,0.735)\end{tabular} \\
			\hline
			\texttt{XGBoost}  & \begin{tabular}{@{}c@{}}0.756 \\ (0.797,0.705)\end{tabular} & \begin{tabular}{@{}c@{}}0.790 \\ (0.865,0.705)\end{tabular} & \begin{tabular}{@{}c@{}}0.792 \\ (0.833,0.714)\end{tabular} & \begin{tabular}{@{}c@{}}0.794 \\ (0.843,0.746)\end{tabular} & \begin{tabular}{@{}c@{}}0.791 \\ (0.833,0.738)\end{tabular}   \\
			\hline
		\end{tabular}
		\caption{Mean of AUPRC (with maximum AUPRC and minimum AUPRC in the bracket) for each number of clusters and each machine learning algorithm}
		\label{auprctable}
	\end{center}
\end{table}

\begin{table}[!ht]
	\begin{center}
		\begin{tabular}{p{17.5mm}|p{21mm}|p{21mm}|p{21mm}|p{21mm}|p{21mm}}
			\hline
			\# clusters & 5 & 6 & 7 & 8 & 9 \\
			\hline
			\texttt{Catboost} & \begin{tabular}{@{}c@{}}0.716 \\ (0.768,0.665)\end{tabular} & \begin{tabular}{@{}c@{}}0.735 \\ (0.777,0.674)\end{tabular} & \begin{tabular}{@{}c@{}}0.715 \\ (0.766,0.669)\end{tabular} & \begin{tabular}{@{}c@{}}0.716 \\ (0.773,0.639)\end{tabular} & \begin{tabular}{@{}c@{}}0.724 \\ (0.767,0.671)\end{tabular}\\
			\hline
			\texttt{LightGBM}   & \begin{tabular}{@{}c@{}}0.689 \\ (0.734,0.670)\end{tabular} & \begin{tabular}{@{}c@{}}0.730 \\ (0.768,0.685)\end{tabular} & \begin{tabular}{@{}c@{}}0.727 \\ (0.775,0.674)\end{tabular} & \begin{tabular}{@{}c@{}}0.716 \\ (0.764,0.656)\end{tabular} & \begin{tabular}{@{}c@{}}0.726 \\ (0.774,0.682)\end{tabular} \\
			\hline
			\texttt{XGBoost}  & \begin{tabular}{@{}c@{}}0.695 \\ (0.738,0.650)\end{tabular} & \begin{tabular}{@{}c@{}}0.726 \\ (0.795,0.643)\end{tabular} & \begin{tabular}{@{}c@{}}0.723 \\ (0.771,0.645)\end{tabular} & \begin{tabular}{@{}c@{}}0.729 \\ (0.781,0.685)\end{tabular} & \begin{tabular}{@{}c@{}}0.723 \\ (0.773,0.656)\end{tabular}   \\
			\hline
		\end{tabular}
		\caption{Mean of Min(Re,P+) (with maximum Min(Re,P+) and minimum Min(Re,P+) in the bracket) for each number of clusters and each machine learning algorithm}
		\label{minreprtable}
	\end{center}
\end{table}

\begin{figure}[!htb]
	\begin{center}
            \includegraphics[width=.48\linewidth]{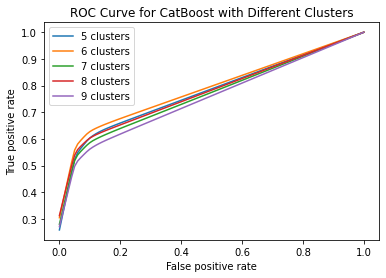}
		\includegraphics[width=.48\linewidth]{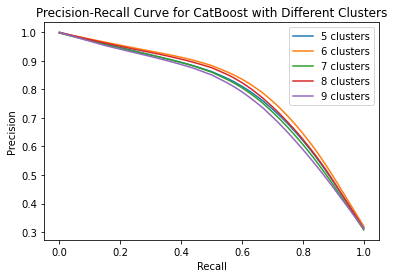}
		\\
            \includegraphics[width=.48\linewidth]{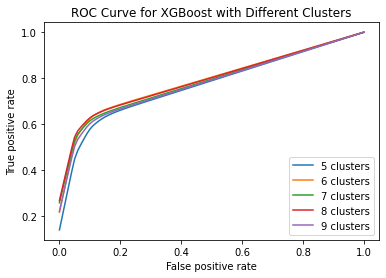}
            \includegraphics[width=.48\linewidth]{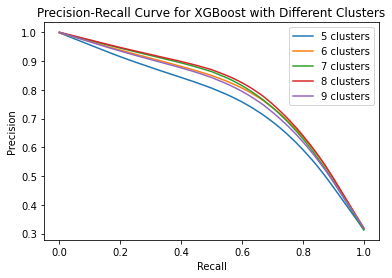}
		\\
		\includegraphics[width=.48\linewidth]{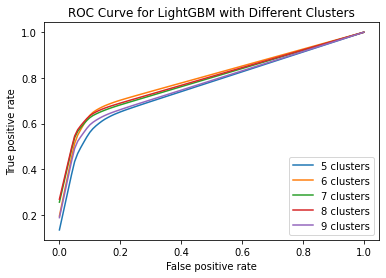}
            \includegraphics[width=.48\linewidth]{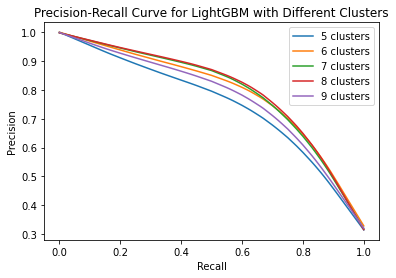}
		\caption{ROC (left) and Precision-Recall Curve (right) for different classifiers with different number of clusters}
		\label{fig:sMRInonzero1}
	\end{center}
\end{figure}

\texttt{Catboost} with 6 clusters provided the best accuracy, AUPRC, and Min(Re, P+). The ROC and precision-recall curves of each classifier and each choice of number of clusters are given in Figure \ref{fig:sMRInonzero1}.

\begin{figure}[!htb]
	\minipage{0.5\textwidth}
	\includegraphics[width=\linewidth]{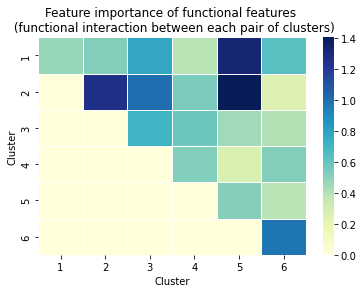}
	\endminipage \hspace{1mm}  \hfill
	\minipage{0.5\textwidth}
	\includegraphics[width=\linewidth]{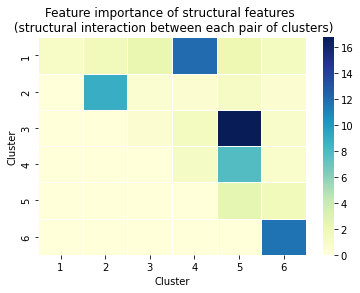}
	\endminipage 
	\caption{Feature importance heatmap for CatBoost with 6 clusters. Each cell of the heatmap corresponds to an entry of the functional or structural connectivity matrix. The darker the color, the higher the level of importance of the entry.}
	\label{fig:heatmap}
\end{figure}

To investigate the features, we provide the feature importance.  The importance of a feature is measured by the change of loss function used in classifier training after removing this feature from the input. Table \ref{tab:feature-importance} gives the 10 most important features among all $6(6+1) + 4=46$ features (as \ref{sec:method-supervised-learning} explained) for \texttt{CatBoost} with 6 Clusters. It shows that structural features are more important than functional features in classification; among demographics and HIV status, age is the most important predictor (the importance rank of race, gender, and HIV status are 24, 45, and 46 respectively). We provide heatmaps (Figure \ref{fig:heatmap}) of feature importance for the fMRI and dMRI features with 6 clusters respectively. Moreover, the structural interaction between the temporal/Default Mode Network (DMN) region and the visual region (occipital lobe) is the most important among all features from the MRI connectome data.

\begin{table}[ht]
	\centering
	\begin{tabular}{rlllll}
		\hline
		Importance Rank & 1 & 2 & 3 & 4 & 5 \\
		\hline
		Feature & str(3,5) & str(1,4) & str(6,6) & str(2,2) & age \\
		\hline
	\end{tabular}
	\caption{Top 5 most important features for CatBoost with 6 Clusters. ``str$(i,j)$'' refers to the entry $(i, j)$ in the structural connectivity average matrix.}
	\label{tab:feature-importance}
\end{table}

\subsection{Comparison with Other Benchmark Methods}

We compare our method in Section \ref{sec:method} with other benchmark methods. First, we claim that our model has advantages over the model using PCA \cite{abdi2010principal}. For each sample, we vectorize dMRI and fMRI matrices after removing redundancy (because they are symmetric) to get an MRI vector and then combine this MRI vector with the demographic features and HIV status. As a result, the data is reformatted as an $n\times p$ matrix, where $n = 275$ is the number of samples and $p = 60766$ is the number of features. Then, we apply PCA to take the first 10 principal components. Finally, we apply the supervised learning algorithms described in Section \ref{sec:method-supervised-learning} to evaluate their performance based on the prediction accuracy. Among those 3 algorithms, the best average prediction accuracy is 0.826 obtained by \texttt{Catboost}, with maximum 0.847 and minimum 0.804, which is lower than the prediction accuracy yielded in our model in Section \ref{sec:method}. Further notice that using principal components weakens the interpretability, as we do not know the meaning of a linear combination of features (ROIs). In contrast, our model with the discrete block structure has high predictability thanks to the membership matrices \cite{Han20}. 

Second, we show that our model has advantages over the model without dimension reduction. We train the 3 supervised learning algorithms in Section \ref{sec:method-supervised-learning}. As we directly input all the 60766 features into the classifier, it takes hours to finish the cross-validation procedure mentioned in Section \ref{sec:method-supervised-learning} so that we cannot reasonably repeat the procedure 100 times and take the average as we did elsewhere. Thus, we only perform the cross-validation once for each of the 3 algorithms and among those 3 algorithms, the best average prediction accuracy is 0.858 obtained by \texttt{XGBoost}, which is similar to the prediction accuracy yielded in our model. However, the time cost is roughly 2 hours, while our model only takes less than 3 minutes. 

Finally, we show the advantage of our clustering result by comparing our accuracy to the accuracy obtained by collecting information utilizing the Brainnetome atlas instead of clustering result obtained by unsupervised learning algorithms. In particular, one can classify 246 regions in the Brainnetome atlas into seven cortical lobes, and subcortical regions (frontal lobe, temporal lobe, parietal lobe, insular lobe, limbic lobe, occipital lobe, and subcortical nuclei). If we use these clusters and apply the supervised machine learning in Section \ref{sec:method-supervised-learning}, the best average prediction accuracy is 0.831 obtained by \texttt{LightGBM}, with maximum 0.851 and minimum 0.807, which is lower than the prediction accuracy yielded in our model in Section \ref{sec:method}. This shows the advantage of our clustering method in Section \ref{sec:uml} as a better methodology for the purpose of predicting cocaine use status.

\section{Discussions}\label{discussion}

In this paper, we utilized an innovative machine learning method to demonstrate that multimodal MRI connectivity features can effectively distinguish  people with and without cocaine use. For this, we create a partial high-order spectral clustering algorithm as an unsupervised machine learning algorithm to reduce the feature size and then applied gradient boosting models to predict cocaine use. Our high value of accuracy, AUROC, AUPRC, and Min(Re, P+) suggests that diffusion and functional connectivity can correctly classify individuals who use cocaine.

By incorporating nested cross-validation, our method is robust against over-fitting. Our approach tries different combinations of the number of clusters and hyper-parameters in three machine learning algorithms (\texttt{Catboost}, \texttt{LightGBM}, and \texttt{XGBoost}). As additional evidence, based on our results, the prediction accuracy does not increase when the number of clusters increases. The issue of over-fitting is also a reason that we only attempted gradient-boosting methods. If we use neural network approaches, we would likely run into the trouble of over-fitting based on the relatively small number of elements in the supervised machine learning model ($275 \times (r(r+1)+4)$).

In our application of the unsupervised machine learning algorithm, we only clustered on regions. It is natural to ask whether there is a block structure on samples. Based on the result of clustering on samples, and clustering on both regions and samples, we did not find a block structure for participants. Therefore, we only assign clusters to regions.

Our method lies in the assumption that there is a block structure inside the data, as we mentioned in Section \ref{sec:uml}. This assumption is reasonable because the regions of interest naturally have different functionalities, and can be grouped into clusters \cite{yeo2014estimates}. On the other hand, the results of our method support the assumption by demonstrating high prediction accuracy. 

For the clustering results, we identified neural networks that were most closely associated with each of the clusters. We use 6 clusters as an example. For example, the six clusters we obtained by our method with $r=6$ appear to belong to the following regions and/or networks: motor, frontoparietal, temporal/default mode network, subcortical, visual, and insular/temporal. Previous literature on cocaine use corroborates the role of these networks as being associated with cocaine use. A prominent theory of how cocaine use disorder originates is through neurobiological changes to subcortical regions due to altered dopamine transmission \citep{everitt2016drug}. Insular regions have been implicated in both cocaine craving \citep{naqvi2014insula} and relapse risk \citep{mchugh2013striatal}. The fronto-parietal network is associated with impulsiveness related to cocaine use and relapse risk \citep{bell2014intact}. The default mode network shows aberrant functioning in cocaine users that is thought to produce difficulties in performing cognitively demanding tasks \citep{zhai2022disrupted}. Finally, cocaine use is associated with deficits in the visual cortex thought to result in reduced inhibitory control \citep{howlett2021inhibitory}. 

Some of the features derived from fMRI and dMRI matrices were greater in importance than others. For example, in the case of 6 clusters, the feature representing the connection between Clusters 3 (temporal/default mode network) and 5 (visual) in dMRI data is the most important among all features. These clusters mapped most closely onto the DMN and visual network. Structural connectivity analyses have shown that white matter association pathways connect the DMN to visual processing regions \citep{alves2019improved}. Future research could focus on the connections between these two networks to further investigate their importance to cocaine use. 

In conclusion, we employed a novel tensor-based unsupervised machine learning technique to identify individuals who use cocaine from those that do not. Using multi-modal MRI imaging data, we were able to achieve high predictive accuracy. This technique provides both interpretable outcomes (clusters of ROIs) and good prediction power. Future studies could parse out the specific relevance of the identified clusters to aspects of cocaine use to provide biomarkers for targeted treatment intervention.

\section*{Acknowledgments}

This research was supported by the Duke Quantitative Methods in HIV/AIDS Research, an NIAID funded Research Education Program (R25 AI140495) and by the Duke University Center for AIDS Research (CFAR), an NIH funded program (5P30 AI064518). Anru R. Zhang was supported in part by NSF Grant NSF CAREER-2203741; Ryan Bell, Shana Hall, Kareem Al-Khalil, and Christina Meade were supported in part by NIH Grants DP2 DA-040226, R01 DA-045565, and R01 DA-047149.
\newpage
\bibliographystyle{elsarticle-num-names}
\bibliography{lit}

\newpage
\appendix

\section{Tensor Algebra \& Algorithms}\label{sec:algorithms}

We illustrate the specific step of the algorithm implemented for unsupervised learning step in this paper. Before illustrating the specific steps of the tensor-based methods implemented in this paper, we give a brief preliminary on tensors, i.e., multiway arrays. The order of a tensor is the number of directions, ways, or modes. We denote a $d$-way array $\cX$ of size $p_1\times \dots \times p_d$ as an order-$d$ dimension-$(p_1,\ldots, p_d)$ tensor and write $\cX \in \mathbb{R}^{p_1\times \dots \times p_d}$. Specifically, an order-$1$ tensor is a vector and an order-$2$ tensor is a matrix. We denote the $i$-th entry of a vector $a$ by $a_i$; the $(i,j)$-th element of a matrix $A$ by $a_{ij}$, the $(i_1,\dots,i_d)$-th element of an order-$d$ tensor $\cX$ by $\cX_{i_1,\dots,i_d}$. For any subsets $\Omega_1, \ldots, \Omega_d$, we use $\cX_{[\Omega_1, \ldots, \Omega_d]}$ to denote the sub-tensor of $\cX$ with $k$-mode indices $\Omega_k$. For the convenience of presentation, we let ``:" alone represent the whole index set. Therefore, $\cX_{[:, :, :, i]}$ means an order-3 tensor acquired from $\cX$ with the mode-4 index fixed to $i$. 

Matricization, also known as unfolding or flattening, is an operator transforming a tensor into a matrix. In particular, the mode-$i$ matricization of $\cX \in \mathbb{R}^{p_1 \times \dots \times p_d}$ is denoted $\cM_i(\cX)$ and satisfies
\[\cM_i(\cX)\in \mathbb{R}^{p_i \times p_1 \times \dots \times p_{i-1} \times p_{i+1} \times \dots \times p_d},\]
\[\cM_i(\cX)_{j_i,j_1+p_1(j_2-1)+\dots+p_1\dots p_{i-1}(j_{i+1}-1)+p_1\dots p_{i-1}p_{i+1}(j_{i+2}-1)+\dots+p_1\dots p_{i-1}p_{i+1}\dots p_{d-1}(j_d-1)}=\cX_{j_1,\dots,j_d}.\]
The $n$-mode (matrix) product of a tensor $\cX \in \mathbb{R}^{r_1 \times \dots \times r_d}$ with a matrix $U_i \in R^{p_i \times r_i}$ is denoted by $\cX \times_i U_i$ and is of size $r_1 \times \dots \times r_{i-1} \times p_i \times r_{i+1} \times r_d$. It satisfies
\[ (\cX \times_i U_i)_{j_1,\dots,j_{i-1},k_i,j_{i+1},\dots,j_d} = \sum_{j_i=1}^{r_i} \cX_{j_1,\dots,j_{i-1},j_i,j_{i+1},\dots,j_d}(U_i)_{k_ij_i} \]
For a tensor $\cS \in \mathbb{R}^{r_1 \times \dots \times r_d}$ and matrices $U_k \in \mathbb{R}^{p_k \times r_k}$ for $k=1,\dots,d$, define the multilinear multiplication as
\[ (\cS \times_1 U_1 \times \dots \times_d U_d)_{i_1,\dots,i_d} = \sum_{j_1=1}^{r_1} \dots \sum_{j_d=1}^{r_d} \cS_{j_1,\dots,j_d} (U_1)_{i_1 j_1} \dots (U_d)_{i_d j_d}. \]
The result $\cS \times_1 U_1 \times \dots \times_d U_d$ is a tensor of shape $(p_1,\dots,p_d)$.
For an array $(r_1,\dots,r_d)$ and $k \in \{1,\dots,d\}$, denote $r_{-k}=\prod_{\substack{1 \le k' \le d \\ k'!= k}} r_{k'}$. 

Next, we provide the pseudocode for the high-order Lloyd in Algorithm \ref{alg:hlloyd} and partial high-order spectral clustering algorithms in Algorithm \ref{alg:phsc}. The two algorithms form the unsupervised learning step to analyze our parsed dataset.
\begin{algorithm}
	\caption{High-order Lloyd Algorithm for Multimodal Connectome Data}\label{alg:hlloyd}
	\begin{algorithmic}
		\Require Connectome dataset $\cY \in \mathbb{R}^{275\times 2\times 246\times 246}$, initialization labels for ROI clusters $z^{(0)} \in \{1,\dots,r\}^{246}$, number of ROI clusters $r$, number of iterations $T$
		\For{$t=0:T-1$}
		\State Update the block means $\cS^{(t)}\in \mathbb{R}^{275\times 2\times r\times r}$ via
		\[ \cS_{i_1,i_2,i_3,i_4}^{(t)} = \mathrm{Average}\left( \left\{ \cY_{i_1,i_2,j_3,j_4}: (z^{(t)})_{j_3}=i_3, (z^{(t)})_{j_4}=i_4 \right\} \right). \]
		\For{$j=1:246$}
		\State Calculate $\cY_4^{(t)} \in \mathbb{R}^{275 \times 2\times246 \times r}$: for each $i_1=1,\ldots, 275; i_2=1,2; i_3=1,\ldots, 246; i_4 = 1,\ldots, r$,
		\[
		(\cY_4^{(t)})_{i_1,i_2,i_3, j} = \mathrm{Average}\left( \left\{ \cY_{i_1,i_2,i_3, j}: (z^{(t)})_{j}=i_4 \right\} \right). 
		\]
		\State Update the membership labels for the $j$-th ROI from $(z^{(t)})_j$ to $(z^{(t+1)})_j$ via
		\[(z^{(t+1)})_j = \underset{a \in \{1,\dots,r\}}{\mathrm{argmin}} \left\| (\cY_4^{(t)})_{[:, :, j, :]} - (\cS^{(t)})_{[:, :, a, :]} \right\|_2^2. \]
		\EndFor
		\EndFor
		\State \Return Estimated block memberships $z^{(T)} \in \{1,\dots,r\}^{246}$
	\end{algorithmic}
\end{algorithm}

\begin{algorithm}
	\caption{Partial High-order Spectral Clustering for Multimodal Connectome Data}\label{alg:phsc}
	\begin{algorithmic}
		\Require Connectome dataset $\cY \in \mathbb{R}^{275 \times 2 \times 246 \times 246}$, number of ROI clusters $r$ 
		\State Compute the singular subspace $\tilde{U}=\mathrm{SVD}_{r}(\cM_{3}(\cY)) \in \mathbb{R}^{246\times r}$.
		\State Compute the singular subspace $\hat{U} = \mathrm{SVD}_{r}(\cM_{3}(\cY \times_{4} \tilde{U}^\top) \in \mathbb{R}^{246\times r}$.
		\State Calculate $\hat{Y}_{3}=\hat{U} \hat{U}^\top \cM_{3}(\cY \times_{4} \hat{U}) \in \mathbb{R}^{246 \times (275\cdot 2\cdot r)}$.
		\State Apply k-means++ to cluster the 246 columns of $\hat{Y}_{3}$ into $r$ clusters. Denote the resulting cluster labels as $z^{(0)} \in \{1,\dots,r\}^{246}$. 
		\State \Return Initialization of memberships $z^{(0)}$
	\end{algorithmic}
\end{algorithm}

\section{More Tables and Figures}

\begin{table}[H]
	\centering
	\begin{tabular}{p{6cm}lllllllllll}
		&&\multicolumn{8}{c}{{Number of Clusters}}\\
		{Region of Interest} & ~ & 5 & 6 & 7 & 8 & 9 & 10 & 11 & 12 & 13 & 14\\ \hline\hline
		\multicolumn{1}{c}{\textbf{Frontal Lobe}} & ~ & ~ & ~ & ~ & ~ & ~ & ~ & ~ & ~ & ~ & ~ \\ 
		\multicolumn{1}{c}{{SFG, Superior Frontal Gyrus}}   & ~ & ~ & ~ & ~ & ~ & ~ & ~ & ~ & ~ & ~ & ~ \\ 
		A8m, medial area 8 & R & 1 & 2 & 1 & 4 & 6 & 1 & 1 & 2 & 1 & 10 \\ 
		A8m, medial area 8 & L & 1 & 2 & 1 & 4 & 6 & 1 & 1 & 2 & 1 & 10 \\ 
		A8dl, dorsolateral area 8 & R & 1 & 2 & 5 & 4 & 5 & 9 & 11 & 2 & 7 & 13 \\ 
		A8dl, dorsolateral area 8 & L & 1 & 2 & 5 & 4 & 5 & 9 & 11 & 2 & 7 & 13 \\ 
		A9l, lateral area 9 & R & 1 & 2 & 5 & 4 & 5 & 9 & 11 & 2 & 7 &  13\\ 
		A9l, lateral area 9 & L & 1 & 2 & 5 & 4 & 5 & 9 & 11 & 2 & 7 & 13\\ 
		A6dl, dorsolateral area 6 & R & 3 & 1 & 2 & 6 & 8 & 8 & 2 & 6 & 9 & 1 \\ 
		A6dl, dorsolateral area 6 & L & 3 & 1 & 2 & 6 & 9 & 8 & 10 & 3 & 3 & 7 \\ 
		A6m, medial area 6 & R & 3 & 1 & 2 & 6 & 8 & 8 & 5 & 3 & 3 & 7 \\ 
		A6m, medial area 6 & L & 3 & 1 & 2 & 6 & 8 & 8 & 5 & 3 & 3 & 7 \\ 
		A9m,medial area 9 & R & 1 & 2 & 5 & 4 & 5 & 9 & 11 & 2 & 7 & 13 \\ 
		A9m,medial area 9 & L & 1 & 2 & 5 & 4 & 5 & 9 & 11 & 2 & 7 & 13\\ 
		A10m, medial area 10 & R & 1 & 2 & 5 & 4 & 5 & 9 & 11 & 2 & 7 & 13 \\ 
		A10m, medial area 10 & L & 1 & 2 & 5 & 4 & 5 & 9 & 11 & 2 & 7 & 13 \\ 
		\multicolumn{1}{c}{{MFG, Middle Frontal Gyrus}}  & ~ & ~ & ~ & ~ & ~ & ~ & ~ & ~ & ~ &  & \\ 
		A9/46d, dorsal area 9/46 & R & 1 & 2 & 5 & 4 & 6 & 9 & 11 & 2 & 7 & 13 \\ 
		A9/46d, dorsal area 9/46 & L & 1 & 2 & 5 & 4 & 6 & 1 & 1 & 11 & 7 & 13 \\ 
		IFJ, inferior frontal junction & R & 1 & 2 & 1 & 4 & 6 & 1 & 1 & 11 & 7 & 13 \\ 
		IFJ, inferior frontal junction & L & 1 & 2 & 1 & 3 & 6 & 1 & 1 & 11 & 1 & 10 \\ 
		A46, area 46 & R & 1 & 2 & 5 & 4 & 5 & 9 & 11 & 2 & 7 & 13 \\ 
		A46, area 46 & L & 1 & 2 & 5 & 4 & 6 & 1 & 1 & 2 & 7 & 13 \\ 
		A9/46v, ventral area 9/46  & R & 1 & 2 & 1 & 3 & 6 & 1 & 1 & 11 & 1 & 10 \\ 
		A9/46v, ventral area 9/46  & L & 1 & 2 & 1 & 3 & 6 & 1 & 1 & 11 & 1 & 10 \\ 
		A8vl, ventrolateral area 8 & R & 1 & 2 & 5 & 4 & 5 & 9 & 11 & 2 & 7 & 13 \\ 
		A8vl, ventrolateral area 8 & L & 1 & 2 & 5 & 4 & 6 & 1 & 1 & 11 & 7 & 13 \\ 
		A6vl, ventrolateral area 6 & R & 1 & 2 & 1 & 3 & 6 & 1 & 1 & 11 & 1 & 10 \\ 
		A6vl, ventrolateral area 6 & L & 1 & 2 & 1 & 3 & 6 & 1 & 1 & 11 & 1 & 10 \\ 
		A10l, lateral area10 & R & 1 & 2 & 5 & 4 & 5 & 9 & 11 & 2 & 7 & 13 \\ 
		A10l, lateral area10 & L & 1 & 2 & 5 & 4 & 5 & 9 & 11 & 2 & 7 & 13 \\ 
	\end{tabular}\caption{List of Clusters}\label{table_clustering_result}
\end{table}
\begin{table}[!]
	\centering
	\begin{tabular}{p{6cm}lllllllllll}
		&&\multicolumn{8}{c}{{Number of Clusters}}\\
		{Region of Interest} & ~ & 5 & 6 & 7 & 8 & 9 & 10 & 11 & 12 & 13 & 14\\ \hline\hline
        \multicolumn{1}{c}{{IFG, Inferior Frontal Gyrus}} & ~ & ~ & ~ & ~ & ~ & ~ & ~ & ~ & ~ &&\\ 
		A44d,dorsal area 44 & R & 1 & 2 & 1 & 3 & 6 & 1 & 1 & 11 & 1 & 10\\ 
		A44d,dorsal area 44 & L & 1 & 2 & 1 & 3 & 6 & 1 & 1 & 11 & 1 & 10\\ 
		IFS, inferior frontal sulcus & R & 1 & 2 & 1 & 3 & 6 & 1 & 1 & 11 & 1 & 10\\ 
		IFS, inferior frontal sulcus & L & 1 & 2 & 1 & 3 & 6 & 1 & 1 & 11 & 1 & 10\\ 
		A45c, caudal area 45 & R & 1 & 2 & 1 & 4 & 6 & 1 & 1 & 11 & 1 & 10\\ 
		A45c, caudal area 45 & L & 1 & 2 & 1 & 3 & 6 & 1 & 1 & 11 & 1 & 10\\ 
		A45r, rostral area 45 & R & 1 & 2 & 1 & 4 & 6 & 1 & 1 & 11 & 1 & 10\\ 
		A45r, rostral area 45 & L & 1 & 2 & 1 & 3 & 6 & 1 & 1 & 11 & 1 & 10\\ 
		A44op, opercular area 44 & R & 1 & 2 & 1 & 4 & 6 & 1 & 1 & 2 & 1 & 10\\ 
		A44op, opercular area 44 & L & 1 & 2 & 1 & 2 & 6 & 1 & 1 & 11 & 1 & 10\\ 
		A44v, ventral area 44 & R & 1 & 6 & 1 & 2 & 6 & 1 & 1 & 9 & 1 & 10\\ 
		A44v, ventral area 44 & L & 1 & 6 & 1 & 3 & 2 & 1 & 1 & 9 & 1 & 10\\ 
		\multicolumn{1}{c}{{OrG, Orbital Gyrus}} & ~ & ~ & ~ & ~ & ~ & ~ & ~ & ~ & ~ && \\ 
		A14m, medial area 14 & R & 2 & 3 & 5 & 4 & 5 & 9 & 11 & 2 & 7 & 2 \\ 
		A14m, medial area 14 & L & 2 & 3 & 5 & 4 & 5 & 9 & 11 & 2 & 7 & 2 \\ 
		A12/47o, orbital area 12/47 & R & 1 & 2 & 7 & 4 & 3 & 3 & 7 & 2 & 7 & 2 \\ 
		A12/47o, orbital area 12/47 & L & 1 & 2 & 1 & 4 & 6 & 1 & 1 & 11 & 1 & 2 \\ 
		A11l, lateral area 11 & R & 2 & 3 & 7 & 4 & 3 & 3 & 7 & 2 & 7 & 2 \\ 
		A11l, lateral area 11 & L & 2 & 3 & 7 & 4 & 3 & 3 & 7 & 2 & 1 & 2 \\ 
		A11m, medial area 11 & R & 2 & 3 & 5 & 4 & 5 & 9 & 11 & 2 & 7 & 2 \\ 
		A11m, medial area 11 & L & 2 & 3 & 5 & 4 & 5 & 9 & 11 & 2 & 7 & 2 \\ 
		A13, area 13 & R & 2 & 3 & 7 & 8 & 3 & 3 & 7 & 4 & 5 & 2 \\ 
		A13, area 13 & L & 2 & 3 & 7 & 8 & 3 & 3 & 7 & 4 & 5 & 2 \\ 
		A12/47l, lateral area 12/47 & R & 1 & 2 & 1 & 4 & 6 & 1 & 1 & 2 & 1 &2 \\ 
		A12/47l, lateral area 12/47 & L & 1 & 2 & 1 & 4 & 6 & 1 & 1 & 11 & 1 &10\\ 
		
	\end{tabular}\caption{List of Clusters, continued}
\end{table}
\begin{table}[!]
	\centering
        \hspace*{-1cm}
	\begin{tabular}{p{8cm}lllllllllll}
		&&\multicolumn{8}{c}{{Number of Clusters}}\\
		{Region of Interest} & ~ & 5 & 6 & 7 & 8 & 9 & 10 & 11 & 12 & 13 & 14\\ \hline\hline
        \multicolumn{1}{c}{{PrG, Precentral Gyrus}} & ~ & ~ & ~ & ~ & ~ & ~ & ~ & ~ & ~ &&\\ 
		A4hf, area 4(head and face region) & R & 3 & 1 & 2 & 6 & 8 & 8 & 5 & 9 & 3 & 7\\ 
		A4hf, area 4(head and face region) & L & 3 & 1 & 2 & 6 & 8 & 8 & 5 & 9 & 9 & 7\\ 
		A6cdl, caudal dorsolateral area 6 & R & 3 & 1 & 2 & 6 & 8 & 8 & 2 & 6 & 3 & 7\\ 
		A6cdl, caudal dorsolateral area 6 & L & 3 & 1 & 2 & 6 & 8 & 8 & 5 & 3 & 3 & 7\\ 
		A4ul, area 4(upper limb region) & R & 3 & 1 & 2 & 6 & 8 & 8 & 2 & 6 & 9 & 1\\ 
		A4ul, area 4(upper limb region) & L & 3 & 1 & 2 & 6 & 9 & 8 & 10 & 3 & 3 & 7\\ 
		A4t, area 4(trunk region) & R & 3 & 1 & 2 & 6 & 8 & 8 & 2 & 6 & 9 & 1\\ 
		A4t, area 4(trunk region) & L & 3 & 1 & 2 & 6 & 9 & 6 & 10 & 10 & 11 & 12\\ 
		A4tl, area 4(tongue and larynx region) & R & 3 & 6 & 4 & 2 & 2 & 2 & 5 & 9 & 10 & 8\\ 
		A4tl, area 4(tongue and larynx region) & L & 3 & 6 & 4 & 2 & 2 & 2 & 5 & 9 & 10 & 8\\ 
		A6cvl, caudal ventrolateral area 6 & R & 3 & 6 & 1 & 3 & 2 & 2 & 5 & 9 & 10 & 10\\ 
		A6cvl, caudal ventrolateral area 6 & L & 3 & 6 & 1 & 3 & 2 & 2 & 5 & 9 & 10 & 10\\ 
        \multicolumn{1}{c}{{PCL, Paracentral Lobule}} & ~ & ~ & ~ & ~ & ~ & ~ & ~ & ~ & ~ &&\\ 
		A1/2/3ll, area1/2/3 (lower limb region) & R & 3 & 1 & 2 & 6 & 8 & 8 & 5 & 3 & 3 & 7 \\ 
		A1/2/3ll, area1/2/3 (lower limb region) & L & 3 & 1 & 2 & 6 & 8 & 8 & 5 & 3 & 3 & 7 \\ 
		A4ll, area 4, (lower limb region) & R & 3 & 1 & 2 & 6 & 8 & 8 & 5 & 3 & 3 & 7 \\ 
		A4ll, area 4, (lower limb region) & L & 3 & 1 & 2 & 6 & 8 & 8 & 5 & 3 & 3 & 7 \\
		\multicolumn{1}{c}{\textbf{Temporal Lobe}} & ~ & ~ & ~ & ~ & ~ & ~ & ~ & ~ & ~ &&\\ 
		\multicolumn{1}{c}{{STG, Superior Temporal Gyrus}} & ~ & ~ & ~ & ~ & ~ & ~ & ~ & ~ & ~ &&\\ 
		A38m, medial area 38 & R & 2 & 3 & 7 & 8 & 3 & 3 & 7 & 4 & 5 & 9 \\ 
		A38m, medial area 38 & L & 2 & 3 & 7 & 8 & 3 & 3 & 7 & 4 & 5 & 9 \\ 
		A41/42, area 41/42 & R & 3 & 6 & 4 & 2 & 2 & 2 & 5 & 9 & 10 & 8 \\ 
		A41/42, area 41/42 & L & 3 & 6 & 4 & 2 & 2 & 2 & 5 & 9 & 10 & 8 \\ 
		TE1.0 and TE1.2 & R & 3 & 6 & 4 & 2 & 2 & 2 & 5 & 9 &10 & 8 \\ 
		TE1.0 and TE1.2 & L & 3 & 6 & 4 & 2 & 2 & 2 & 5 & 9 &10 & 8 \\ 
		A22c, caudal area 22 & R & 3 & 6 & 4 & 2 & 2 & 2 & 5 & 9 &10 & 8 \\ 
		A22c, caudal area 22 & L & 3 & 6 & 4 & 2 & 2 & 2 & 5 & 9 &10 & 8 \\ 
		A38l, lateral area 38 & R & 2 & 3 & 7 & 8 & 3 & 3 & 7 & 4 & 5 & 9 \\ 
		A38l, lateral area 38 & L & 2 & 3 & 7 & 8 & 3 & 3 & 7 & 4 & 5 & 9 \\ 
		A22r, rostral area 22 & R & 3 & 6 & 4 & 2 & 2 & 2 & 5 & 9 &10 & 8 \\ 
		A22r, rostral area 22 & L & 3 & 6 & 4 & 2 & 2 & 2 & 5 & 9 &10 & 8 \\ 
	\end{tabular}\caption{List of Clusters, continued}
\end{table}
\begin{table}[!]
	\centering
        \hspace*{-1cm}
	\begin{tabular}{p{8cm}lllllllllll}
		&&\multicolumn{8}{c}{{Number of Clusters}}\\
		{Region of Interest} & ~ & 5 & 6 & 7 & 8 & 9 & 10 & 11 & 12 & 13 & 14\\ \hline\hline
        \multicolumn{1}{c}{{MTG, Middle Temporal Gyrus}} & ~ & ~ & ~ & ~ & ~ & ~ & ~ & ~ & ~ &&\\ 
		A21c, caudal area 21 & R & 2 & 3 & 7 & 8 & 3 & 3 & 7 & 4 & 5 & 9 \\ 
		A21c, caudal area 21 & L & 2 & 3 & 7 & 8 & 3 & 3 & 7 & 11 & 5 & 9 \\ 
		A21r, rostral area 21 & R & 2 & 3 & 7 & 8 & 3 & 3 & 7 & 4 & 5 & 9 \\ 
		A21r, rostral area 21 & L & 2 & 3 & 7 & 8 & 3 & 3 & 7 & 4 & 5 & 9 \\ 
		A37dl, dorsolateral area37 & R & 3 & 6 & 4 & 3 & 2 & 2 & 9 & 9 &10 & 8\\ 
		A37dl, dorsolateral area37 & L & 3 & 6 & 4 & 3 & 2 & 2 & 9 & 9 &10 & 3\\ 
		aSTS, anterior superior temporal sulcus & R & 2 & 3 & 7 & 8 & 3 & 3 & 7 & 4 & 5 & 9 \\ 
		aSTS, anterior superior temporal sulcus & L & 2 & 3 & 7 & 8 & 3 & 3 & 7 & 4 & 5 & 9 \\ 
        \multicolumn{1}{c}{{ITG, Inferior Temporal Gyrus}} & ~ & ~ & ~ & ~ & ~ & ~ & ~ & ~ & ~ &&\\ 
		A20iv, intermediate ventral area 20 & R & 2 & 3 & 7 & 8 & 3 & 3 & 7 & 4 & 5 & 9 \\ 
		A20iv, intermediate ventral area 20 & L & 2 & 3 & 7 & 8 & 3 & 3 & 7 & 4 & 5 & 9 \\ 
		A37elv, extreme lateroventral area37 & R & 5 & 5 & 3 & 3 & 4 & 7 & 9 & 11& 12 & 3 \\ 
		A37elv, extreme lateroventral area37 & L & 5 & 5 & 3 & 3 & 4 & 7 & 9 & 11& 12 & 3 \\ 
		A20r, rostral area 20 & R & 2 & 3 & 7 & 8 & 3 & 3 & 7 & 4 & 5 & 9 \\ 
		A20r, rostral area 20 & L & 2 & 3 & 7 & 8 & 3 & 3 & 7 & 4 & 5 & 9 \\ 
		A20il, intermediate lateral area 20 & R & 2 & 3 & 7 & 8 & 3 & 3 & 7 & 4 & 5 & 9 \\ 
		A20il, intermediate lateral area 20 & L & 2 & 3 & 7 & 8 & 3 & 3 & 7 & 4 & 5 & 9 \\ 
		A37vl, ventrolateral area 37 & R & 5 & 5 & 1 & 3 & 4 & 7 & 9 & 11 &12 & 3 \\ 
		A37vl, ventrolateral area 37 & L & 5 & 5 & 3 & 3 & 4 & 7 & 9 & 9 &12 & 3 \\ 
		A20cl, caudolateral of area 20 & R & 2 & 3 & 7 & 4 & 3 & 3 & 7 & 11 & 7 & 9 \\ 
		A20cl, caudolateral of area 20 & L & 2 & 3 & 7 & 3 & 3 & 3 & 7 & 11 & 5 & 9 \\ 
		A20cv, caudoventral of area 20 & R & 2 & 3 & 7 & 8 & 3 & 3 & 7 & 11 & 5 & 9 \\ 
		A20cv, caudoventral of area 20 & L & 2 & 3 & 7 & 3 & 3 & 3 & 7 & 11 & 5 & 9 \\ 
		\multicolumn{1}{c}{{FuG, Fusiform Gyrus}} & ~ & ~ & ~ & ~ & ~ & ~ & ~ & ~ & ~ &&\\ 
		A20rv, rostroventral area 20 & R & 2 & 3 & 7 & 8 & 3 & 3 & 7 & 4 & 5 & 9 \\ 
		A20rv, rostroventral area 20 & L & 2 & 3 & 7 & 8 & 3 & 3 & 7 & 4 & 5 & 9 \\ 
		A37mv, medioventral area37 & R & 5 & 5 & 3 & 5 & 4 & 5 & 4 & 5 & 2 & 4\\ 
		A37mv, medioventral area37 & L & 5 & 5 & 3 & 5 & 4 & 5 & 4 & 5 & 2 & 4\\ 
		A37lv, lateroventral area37 & R & 5 & 5 & 3 & 5 & 4 & 7 & 9 & 5 & 12 & 3 \\ 
		A37lv, lateroventral area37 & L & 5 & 5 & 3 & 5 & 4 & 7 & 9 & 5 & 12 & 3 \\ 
	\end{tabular}\caption{List of Clusters, continued}
\end{table}
\begin{table}[!]
	\centering
        \hspace*{-2.5cm}
	\begin{tabular}{p{10cm}lllllllllll}
		&&\multicolumn{8}{c}{{Number of Clusters}}\\
		{Region of Interest} & ~ & 5 & 6 & 7 & 8 & 9 & 10 & 11 & 12 & 13 & 14\\ \hline\hline
        \multicolumn{1}{c}{{PhG, Parahippocampal Gyrus}} & ~ & ~ & ~ & ~ & ~ & ~ & ~ & ~ & ~ \\ 
		A35/36r, rostral area 35/36 & R & 2 & 3 & 7 & 8 & 3 & 3 & 7 & 4 & 5 & 9 \\ 
		A35/36r, rostral area 35/36 & L & 2 & 3 & 7 & 8 & 3 & 3 & 7 & 4 & 5 & 9 \\ 
		A35/36c, caudal area 35/36 & R & 2 & 3 & 7 & 8 & 3 & 3 & 7 & 4 & 5 & 9 \\ 
		A35/36c, caudal area 35/36 & L & 2 & 3 & 7 & 8 & 3 & 3 & 7 & 4 & 5 & 9 \\ 
		TL, area TL (lateral PPHC) & R & 2 & 3 & 7 & 8 & 3 & 3 & 7 & 4 & 5 & 9 \\ 
		TL, area TL (lateral PPHC) & L & 2 & 3 & 7 & 8 & 3 & 3 & 7 & 4 & 5 & 9 \\ 
		A28/34, area 28/34 (EC, entorhinal cortex) & R & 2 & 3 & 7 & 8 & 3 & 3 & 7 & 4 & 5 & 9 \\ 
		A28/34, area 28/34 (EC, entorhinal cortex) & L & 2 & 3 & 7 & 8 & 3 & 3 & 7 & 4 & 5 & 9 \\ 
		TI, area TI(temporal agranular insular cortex) & R & 2 & 3 & 7 & 8 & 3 & 3 & 7 & 4 & 5 & 9 \\ 
		TI, area TI(temporal agranular insular cortex) & L & 2 & 3 & 7 & 8 & 3 & 3 & 7 & 4 & 5 & 9 \\ 
		TH, area TH (medial PPHC) & R & 2 & 3 & 7 & 8 & 3 & 3 & 7 & 4 & 5 & 9 \\ 
		TH, area TH (medial PPHC) & L & 2 & 3 & 7 & 8 & 3 & 3 & 7 & 4 & 5 & 9 \\ 
        \multicolumn{1}{c}{{pSTS, posterior Superior Temporal Sulcus}}  & ~ & ~ & ~ & ~ & ~ & ~ & ~ & ~ & ~ &&\\ 
		rpSTS, rostroposterior superior temporal sulcus & R & 3 & 6 & 4 & 2 & 2 & 2 & 7 & 9 & 10 & 8\\ 
		rpSTS, rostroposterior superior temporal sulcus & L & 3 & 6 & 4 & 2 & 2 & 2 & 5 & 9 & 10 & 8\\ 
		cpSTS, caudoposterior superior temporal sulcus & R & 3 & 6 & 4 & 2 & 2 & 2 & 7 & 9 & 10 & 8\\ 
		cpSTS, caudoposterior superior temporal sulcus & L & 3 & 6 & 4 & 2 & 2 & 2 & 5 & 9 & 10 & 8\\ 
		\multicolumn{1}{c}{\textbf{Parietal Lobe}} & ~ & ~ & ~ & ~ & ~ & ~ & ~ & ~ & ~ &&\\ 
		\multicolumn{1}{c}{{SPL, Superior Parietal Lobule}} & ~ & ~ & ~ & ~ & ~ & ~ & ~ & ~ & ~ &&\\ 
		A7r, rostral area 7 & R & 5 & 5 & 3 & 6 & 4 & 7 & 9 & 3 & 12 & 3 \\ 
		A7r, rostral area 7 & L & 5 & 1 & 2 & 6 & 8 & 8 & 9 & 3 & 12 & 7 \\ 
		A7c, caudal area 7 & R & 5 & 5 & 3 & 5 & 4 & 7 & 9 & 5 & 12 & 3 \\ 
		A7c, caudal area 7 & L & 5 & 5 & 3 & 5 & 4 & 7 & 9 & 5 & 12 & 3 \\ 
		A5l, lateral area 5 & R & 3 & 5 & 3 & 3 & 2 & 7 & 9 & 9 & 12 & 3 \\ 
		A5l, lateral area 5 & L & 3 & 1 & 2 & 6 & 8 & 8 & 9 & 3 & 3 & 7\\ 
		A7pc, postcentral area 7 & R & 3 & 1 & 2 & 6 & 8 & 8 & 2 & 6 & 9 & 1\\ 
		A7pc, postcentral area 7 & L & 3 & 1 & 2 & 6 & 9 & 6 & 10 & 10 & 11 & 12\\ 
		A7ip, intraparietal area 7(hIP3) & R & 5 & 5 & 3 & 3 & 4 & 7 & 9 & 5 & 12 & 3\\ 
		A7ip, intraparietal area 7(hIP3) & L & 5 & 5 & 3 & 3 & 4 & 7 & 9 & 3 & 12 & 3\\ 
	\end{tabular}\caption{List of Clusters, continued}
\end{table}
\begin{table}[!]
	\centering
        \hspace*{-1cm}
	\begin{tabular}{p{8cm}lllllllllll}
		&&\multicolumn{8}{c}{{Number of Clusters}}\\
		{Region of Interest} & ~ & 5 & 6 & 7 & 8 & 9 & 10 & 11 & 12 & 13 & 14\\ \hline\hline
        \multicolumn{1}{c}{{IPL, Inferior Parietal Lobule}} & ~ & ~ & ~ & ~ & ~ & ~ & ~ & ~ & ~ &&\\ 
		A39c, caudal area 39(PGp) & R & 5 & 5 & 3 & 5 & 4 & 7 & 9 & 11 & 7 & 3 \\ 
		A39c, caudal area 39(PGp) & L & 5 & 5 & 3 & 5 & 4 & 7 & 9 & 11 & 12 & 3 \\ 
		A39rd, rostrodorsal area 39(Hip3) & R & 1 & 2 & 5 & 4 & 5 & 9 & 11 & 11 & 7 & 13 \\ 
		A39rd, rostrodorsal area 39(Hip3) & L & 1 & 2 & 5 & 4 & 5 & 1 & 11 & 11 & 7 & 13 \\ 
		A40rd, rostrodorsal area 40(PFt) & R & 3 & 6 & 1 & 3 & 2 & 7 & 9 & 9 & 12 & 3 \\ 
		A40rd, rostrodorsal area 40(PFt) & L & 3 & 6 & 1 & 3 & 2 & 7 & 9 & 9 & 12 & 3 \\ 
		A40c, caudal area 40(PFm) & R & 1 & 2 & 1 & 4 & 6 & 1 & 1 & 11 & 1 & 13 \\ 
		A40c, caudal area 40(PFm) & L & 1 & 2 & 1 & 3 & 6 & 1 & 1 & 11 & 1 & 10 \\ 
		A39rv, rostroventral area 39(PGa) & R & 2 & 3 & 5 & 4 & 5 & 9 & 11 & 2 & 7 & 13 \\ 
		A39rv, rostroventral area 39(PGa) & L & 1 & 2 & 5 & 4 & 5 & 9 & 11 & 11 & 7 & 13 \\ 
		A40rv, rostroventral area 40(PFop) & R & 3 & 6 & 4 & 2 & 2 & 2 & 5 & 9 & 10 & 8\\ 
		A40rv, rostroventral area 40(PFop) & L & 3 & 6 & 4 & 2 & 2 & 2 & 5 & 9 & 10 & 8 \\ 
        \multicolumn{1}{c}{{Pcun, Precuneus}} & ~ & ~ & ~ & ~ & ~ & ~ & ~ & ~ & ~ && \\ 
		A7m, medial area 7(PEp) & R & 5 & 5 & 3 & 5 & 4 & 7 & 9 & 8 & 8 & 14 \\ 
		A7m, medial area 7(PEp) & L & 5 & 5 & 3 & 5 & 4 & 7 & 9 & 8 & 8 & 14 \\ 
		A5m, medial area 5(PEm) & R & 3 & 1 & 2 & 6 & 8 & 8 & 9 & 3 & 12 & 7 \\ 
		A5m, medial area 5(PEm) & L & 3 & 1 & 2 & 6 & 8 & 8 & 9 & 3 & 12 & 7 \\ 
		dmPOS, dorsomedial parietooccipital  sulcus(PEr)  & R & 5 & 5 & 3 & 5 & 4 & 5 & 4 & 8 & 8 & 14 \\ 
		dmPOS, dorsomedial parietooccipital  sulcus(PEr)  & L & 5 & 5 & 3 & 5 & 4 & 5 & 4 & 8 & 8 & 14 \\ 
		A31, area 31 (Lc1) & R & 2 & 3 & 5 & 4 & 5 & 9 & 11 & 8 & 8 & 14 \\ 
		A31, area 31 (Lc1) & L & 2 & 3 & 5 & 4 & 5 & 9 & 11 & 8 & 8 & 14 \\ 
	\end{tabular}\caption{List of Clusters, continued}
\end{table}
\begin{table}[!]
	\centering\hspace*{-1cm}
	\begin{tabular}{p{8cm}lllllllllll}
		&&\multicolumn{8}{c}{{Number of Clusters}}\\
		{Region of Interest} & ~ & 5 & 6 & 7 & 8 & 9 & 10 & 11 & 12 & 13 & 14\\ \hline\hline
        \multicolumn{1}{c}{{PoG, Postcentral Gyrus}} & ~ & ~ & ~ & ~ & ~ & ~ & ~ & ~ & ~ &&\\ 
		A1/2/3ulhf, area 1/2/3(upper limb, head and face region) & R & 3 & 1 & 2 & 6 & 8 & 8 & 5 & 9 & 3 & 7\\ 
		A1/2/3ulhf, area 1/2/3(upper limb, head and face region) & L & 3 & 1 & 2 & 6 & 8 & 8 & 5 & 3 & 3 & 7\\ 
		A1/2/3tonIa, area 1/2/3(tongue and larynx region) & R & 3 & 6 & 4 & 2 & 2 & 2 & 5 & 9 &10 & 8\\ 
		A1/2/3tonIa, area 1/2/3(tongue and larynx region) & L & 3 & 6 & 4 & 2 & 2 & 2 & 5 & 9 &10 & 8\\ 
		A2, area 2 & R & 3 & 1 & 2 & 6 & 8 & 8 & 5 & 9 & 3 & 7\\ 
		A2, area 2 & L & 3 & 1 & 2 & 6 & 8 & 8 & 5 & 3 & 3 & 7\\ 
		A1/2/3tru, area1/2/3(trunk region) & R & 3 & 1 & 2 & 6 & 8 & 8 & 2 & 6 & 9 & 1\\ 
		A1/2/3tru, area1/2/3(trunk region) & L & 3 & 1 & 2 & 6 & 9 & 6 & 10 & 10 & 11 & 12\\ 
        \multicolumn{1}{c}{\textbf{Insular Lobe}} & ~ & ~ & ~ & ~ & ~ & ~ & ~ & ~ & ~ &&\\ 
		\multicolumn{1}{c}{{INS, Insular Gyrus}} & ~ & ~ & ~ & ~ & ~ & ~ & ~ & ~ & ~ &&\\ 
		G, hypergranular insula & R & 3 & 6 & 4 & 2 & 2 & 2 & 5 & 9 & 10 & 8\\ 
		G, hypergranular insula & L & 3 & 6 & 4 & 2 & 2 & 2 & 5 & 9 & 10 & 8\\ 
		vIa, ventral agranular insula & R & 1 & 3 & 7 & 8 & 3 & 3 & 7 & 4 & 1 & 2\\ 
		vIa, ventral agranular insula & L & 1 & 3 & 7 & 8 & 3 & 3 & 7 & 4 & 1 & 2\\ 
		dIa, dorsal agranular insula & R & 1 & 6 & 1 & 2 & 6 & 1 & 1 & 9& 1 &10 \\ 
		dIa, dorsal agranular insula & L & 1 & 6 & 1 & 2 & 6 & 1 & 1 & 9 & 1 & 10\\ 
		vId/vIg, ventral dysgranular and granular insula & R & 3 & 6 & 4 & 2 & 2 & 2 & 7 & 9 & 10 & 8\\ 
		vId/vIg, ventral dysgranular and granular insula & L & 3 & 6 & 4 & 2 & 2 & 2 & 7 & 9 & 10 & 8\\ 
		dIg, dorsal granular insula & R & 3 & 6 & 4 & 2 & 2 & 2 & 5 & 9 & 10 & 8\\ 
		dIg, dorsal granular insula & L & 3 & 6 & 4 & 2 & 2 & 2 & 5 & 9 & 10 & 8\\ 
		dId, dorsal dysgranular insula & R & 3 & 6 & 4 & 2 & 2 & 2 & 5 & 9 & 10 & 8\\ 
		dId, dorsal dysgranular insula & L & 3 & 6 & 4 & 2 & 2 & 2 & 5 & 9 & 10 & 8\\ 
	\end{tabular}
	\caption{List of Clusters, continued}
\end{table}

\begin{table}[!]
	\centering\hspace*{-1cm}
	\begin{tabular}{p{8cm}lllllllllll}
		&&\multicolumn{8}{c}{{Number of Clusters}}\\
		{Region of Interest} & ~ & 5 & 6 & 7 & 8 & 9 & 10 & 11 & 12 & 13 & 14\\ \hline\hline
        \multicolumn{1}{c}{\textbf{Limbic Lobe}} & ~ & ~ & ~ & ~ & ~ & ~ & ~ & ~ & ~ &&\\ 
		\multicolumn{1}{c}{{CG, Cingulate Gyrus}} & ~ & ~ & ~ & ~ & ~ & ~ & ~ & ~ & ~ &&\\ 
		A23d, dorsal area 23 & R & 2 & 3 & 5 & 4 & 5 & 9 & 11 & 8 & 8 & 14\\ 
		A23d, dorsal area 23 & L & 2 & 3 & 5 & 4 & 5 & 9 & 11 & 8 & 8 & 14\\ 
		A24rv, rostroventral area 24 & R & 1 & 2 & 5 & 4 & 6 & 1 & 1 & 2 & 1 & 2\\ 
		A24rv, rostroventral area 24 & L & 1 & 2 & 5 & 4 & 5 & 9 & 11 & 2 & 1 & 2\\ 
		A32p, pregenual area 32 & R & 1 & 2 & 5 & 4 & 5 & 9 & 11 & 2 & 7 & 2\\ 
		A32p, pregenual area 32 & L & 1 & 2 & 5 & 4 & 6 & 1 & 1 & 2 & 1 & 2\\ 
		A23v, ventral area 23 & R & 2 & 3 & 5 & 4 & 5 & 9 & 11 & 8 & 8 & 14\\ 
		A23v, ventral area 23 & L & 2 & 3 & 5 & 4 & 5 & 9 & 11 & 8 & 8 & 14\\ 
		A24cd, caudodorsal area 24 & R & 3 & 6 & 4 & 2 & 2 & 2 & 5 & 9 & 10 & 8 \\ 
		A24cd, caudodorsal area 24 & L & 3 & 6 & 4 & 2 & 2 & 2 & 5 & 9 & 10 & 8 \\ 
		A23c, caudal area 23 & R & 3 & 6 & 4 & 2 & 2 & 2 & 5 & 9 & 10 & 8 \\ 
		A23c, caudal area 23 & L & 3 & 6 & 4 & 2 & 2 & 2 & 5 & 9 & 10 & 8 \\ 
		A32sg, subgenual area 32 & R & 2 & 3 & 5 & 4 & 5 & 9 & 11 & 2 & 7 & 2\\ 
		A32sg, subgenual area 32 & L & 1 & 2 & 5 & 4 & 5 & 9 & 11 & 2 & 7 & 2\\ 	
        \multicolumn{1}{c}{\textbf{Occipital Lobe}} & ~ & ~ & ~ & ~ & ~ & ~ & ~ & ~ & ~ &&\\ 
		\multicolumn{1}{c}{{ MVOcC, MedioVentral Occipital Cortex}} & ~ & ~ & ~ & ~ & ~ & ~ & ~ & ~ & ~ &&\\ 
		cLinG, caudal lingual gyrus & R & 5 & 5 & 3 & 5 & 4 & 5 & 4 & 5 & 2 & 4\\ 
		cLinG, caudal lingual gyrus & L & 5 & 5 & 3 & 5 & 4 & 5 & 4 & 5 & 2 & 4\\ 
		rCunG, rostral cuneus gyrus & R & 5 & 5 & 3 & 5 & 4 & 5 & 4 & 5 & 2 & 4\\ 
		rCunG, rostral cuneus gyrus & L & 5 & 5 & 3 & 5 & 4 & 5 & 4 & 5 & 2 & 4\\ 
		cCunG, caudal cuneus gyrus & R & 5 & 5 & 3 & 5 & 4 & 5 & 4 & 5 & 2 & 4\\ 
		cCunG, caudal cuneus gyrus & L & 5 & 5 & 3 & 5 & 4 & 5 & 4 & 5 & 2 & 4\\ 
		rLinG, rostral lingual gyrus & R & 5 & 5 & 3 & 5 & 4 & 5 & 4 & 5 & 2 & 4\\ 
		rLinG, rostral lingual gyrus & L & 5 & 5 & 3 & 5 & 4 & 5 & 4 & 5 & 2 & 4\\ 
		vmPOS,ventromedial parietooccipital sulcus & R & 5 & 5 & 3 & 5 & 4 & 5 & 4 & 5 & 2 & 4\\ 
		vmPOS,ventromedial parietooccipital sulcus & L & 5 & 5 & 3 & 5 & 4 & 5 & 4 & 5 & 2 & 4\\ 
	\end{tabular}
	\caption{List of Clusters, continued}
\end{table}

\begin{table}[!]
	\centering\hspace*{-1cm}
	\begin{tabular}{p{8cm}lllllllllll}
		&&\multicolumn{8}{c}{{Number of Clusters}}\\
		{Region of Interest} & ~ & 5 & 6 & 7 & 8 & 9 & 10 & 11 & 12 & 13 & 14\\ \hline\hline
        \multicolumn{1}{c}{{LOcC, lateral Occipital Cortex}} & ~ & ~ & ~ & ~ & ~ & ~ & ~ & ~ & ~ \\ 
		mOccG, middle occipital gyrus & R & 5 & 5 & 3 & 5 & 4 & 5 & 4 & 5 & 2 & 4\\ 
		mOccG, middle occipital gyrus & L & 5 & 5 & 3 & 5 & 4 & 5 & 4 & 5 & 2 & 4\\ 
		V5/MT+, area V5/MT+ & R & 5 & 5 & 3 & 5 & 4 & 7 & 9 & 5 & 12 & 3\\ 
		V5/MT+, area V5/MT+ & L & 5 & 5 & 3 & 5 & 4 & 7 & 9 & 5 & 12 & 3\\ 
		OPC, occipital polar cortex & R & 5 & 5 & 3 & 5 & 4 & 5 & 4 & 5 & 2 & 4\\ 
		OPC, occipital polar cortex & L & 5 & 5 & 3 & 5 & 4 & 5 & 4 & 5 & 2 & 4\\ 
		iOccG, inferior occipital gyrus & R & 5 & 5 & 3 & 5 & 4 & 5 & 4 & 5 & 2 & 4\\ 
		iOccG, inferior occipital gyrus & L & 5 & 5 & 3 & 5 & 4 & 5 & 4 & 5 & 2 & 4\\ 
		msOccG, medial superior occipital gyrus & R & 5 & 5 & 3 & 5 & 4 & 5 & 4 & 5 & 2 & 4\\ 
		msOccG, medial superior occipital gyrus & L & 5 & 5 & 3 & 5 & 4 & 5 & 4 & 5 & 2 & 4\\ 
		lsOccG, lateral superior occipital gyrus & R & 5 & 5 & 3 & 5 & 4 & 7 & 9 & 5 & 12 & 3\\ 
		lsOccG, lateral superior occipital gyrus & L & 5 & 5 & 3 & 5 & 4 & 7 & 9 & 5 & 12 & 3\\ 
		\multicolumn{1}{c}{\textbf{Subcortical Nuclei}} & ~ & ~ & ~ & ~ & ~ & ~ & ~ & ~ & ~ &&\\ 
		\multicolumn{1}{c}{{Amyg, Amygdala}} & ~ & ~ & ~ & ~ & ~ & ~ & ~ & ~ & ~ &&\\ 
		mAmyg, medial amygdala & R & 2 & 3 & 7 & 8 & 3 & 3 & 7 & 4 & 5 & 9 \\ 
		mAmyg, medial amygdala & L & 2 & 3 & 7 & 8 & 3 & 3 & 7 & 4 & 5 & 9 \\ 
		lAmyg, lateral amygdala & R & 2 & 3 & 7 & 8 & 3 & 3 & 7 & 4 & 5 & 9 \\ 
		lAmyg, lateral amygdala & L & 2 & 3 & 7 & 8 & 3 & 3 & 7 & 4 & 5 & 9 \\ 
		\multicolumn{1}{c}{{Hipp, Hippocampus}} & ~ & ~ & ~ & ~ & ~ & ~ & ~ & ~ & ~ &&\\ 
		rHipp, rostral hippocampus & R & 2 & 3 & 7 & 8 & 3 & 3 & 7 & 4 & 5 & 9 \\ 
		rHipp, rostral hippocampus & L & 2 & 3 & 7 & 8 & 3 & 3 & 7 & 4 & 5 & 9 \\ 
		cHipp, caudal hippocampus & R & 2 & 3 & 7 & 8 & 3 & 3 & 7 & 4 & 5 & 9 \\ 
		cHipp, caudal hippocampus & L & 2 & 3 & 7 & 8 & 3 & 3 & 7 & 4 & 5 & 9 \\ 
	\end{tabular}
	\caption{List of Clusters, continued}
\end{table}

\begin{table}[!]
	\centering\hspace*{-1cm}
	\begin{tabular}{p{8cm}lllllllllll}
		&&\multicolumn{8}{c}{{Number of Clusters}}\\
		{Region of Interest} & ~ & 5 & 6 & 7 & 8 & 9 & 10 & 11 & 12 & 13 & 14\\ \hline\hline
        \multicolumn{1}{c}{{BG, Basal Ganglia}} & ~ & ~ & ~ & ~ & ~ & ~ & ~ & ~ & ~ &&\\ 
		vCa, ventral caudate & R & 4 & 4 & 6 & 1 & 1 & 4 & 8 & 1 &4 & 6\\ 
		vCa, ventral caudate & L & 4 & 4 & 6 & 7 & 1 & 4 & 8 & 1 &6 & 6\\ 
		GP, globus pallidus & R & 4 & 4 & 6 & 1 & 1 & 4 & 6 & 12 &4 & 11\\ 
		GP, globus pallidus & L & 4 & 4 & 6 & 7 & 7 & 4 & 3 & 1 &6 & 6\\ 
		NAC, nucleus accumbens & R & 4 & 4 & 6 & 1 & 1 & 4 & 8 & 1 &4 & 6\\ 
		NAC, nucleus accumbens & L & 4 & 4 & 6 & 7 & 1 & 4 & 8 & 1 &6 & 6\\ 
		vmPu, ventromedial putamen & R & 4 & 4 & 6 & 1 & 1 & 4 & 8 & 1 &4 & 6\\ 
		vmPu, ventromedial putamen & L & 4 & 4 & 6 & 7 & 1 & 4 & 8 & 1 &6 & 6\\ 
		dCa, dorsal caudate & R & 4 & 4 & 6 & 1 & 1 & 4 & 8 & 1 &4 & 6\\ 
		dCa, dorsal caudate & L & 4 & 4 & 6 & 7 & 1 & 4 & 8 & 1 &6 & 6\\ 
		dlPu, dorsolateral putamen & R & 4 & 4 & 6 & 1 & 1 & 4 & 6 & 12 &4 & 11\\ 
		dlPu, dorsolateral putamen & L & 4 & 4 & 6 & 7 & 7 & 10 & 3 & 7 &6 & 5\\ 
		\multicolumn{1}{c}{{Tha, Thalamus}} & ~ & ~ & ~ & ~ & ~ & ~ & ~ & ~ & ~ &&\\ 
		mPFtha, medial pre-frontal thalamus & R & 4 & 4 & 6 & 1 & 1 & 4 & 8 & 1 &4 & 6\\ 
		mPFtha, medial pre-frontal thalamus & L & 4 & 4 & 6 & 7 & 1 & 4 & 8 & 1 &6 & 6\\ 
		mPMtha, pre-motor thalamus & R & 4 & 4 & 6 & 1 & 1 & 4 & 6 & 12 &4 & 11\\ 
		mPMtha, pre-motor thalamus & L & 4 & 4 & 6 & 7 & 7 & 4 & 3 & 1 &6 & 6\\ 
		Stha, sensory thalamus & R & 4 & 4 & 6 & 1 & 1 & 4 & 6 & 12 &4 & 11\\ 
		Stha, sensory thalamus & L & 4 & 4 & 6 & 7 & 7 & 10 & 3 & 7 &13 & 5\\ 
		rTtha, rostral temporal thalamus & R & 4 & 4 & 6 & 1 & 1 & 4 & 8 & 1 &4 & 6\\ 
		rTtha, rostral temporal thalamus & L & 4 & 4 & 6 & 7 & 1 & 4 & 8 & 1 &6 & 6\\ 
		PPtha, posterior parietal thalamus & R & 4 & 4 & 6 & 1 & 1 & 4 & 8 & 1 &4 & 6\\ 
		PPtha, posterior parietal thalamus & L & 4 & 4 & 6 & 7 & 1 & 4 & 8 & 1 &6 & 6\\ 
		Otha, occipital thalamus & R & 4 & 4 & 6 & 1 & 1 & 4 & 8 & 1 &4 & 6\\ 
		Otha, occipital thalamus & L & 4 & 4 & 6 & 7 & 1 & 4 & 8 & 1 &6 & 6\\ 
		cTtha, caudal temporal thalamus & R & 4 & 4 & 6 & 1 & 1 & 4 & 8 & 1 &4 & 6\\ 
		cTtha, caudal temporal thalamus & L & 4 & 4 & 6 & 7 & 1 & 4 & 8 & 1 &6 & 6\\ 
		lPFtha, lateral pre-frontal thalamus & R & 4 & 4 & 6 & 1 & 1 & 4 & 6 & 12 &4 & 11\\ 
		lPFtha, lateral pre-frontal thalamus & L & 4 & 4 & 6 & 7 & 1 & 4 & 8 & 1 &6 & 6\\ 
	\end{tabular}
	\caption{List of Clusters, continued}
\end{table}







\end{sloppypar}
\end{document}